\documentclass[9pt,twocolumn,twoside]{opticajnl}
\journal{opticajournal} 

\setboolean{shortarticle}{true}

\usepackage{widetext}
\usepackage{caption}
\usepackage[toc,page]{appendix}
\usepackage{braket}
\usepackage{amsmath}
\usepackage{amsfonts}
\usepackage{amssymb}
\usepackage{graphicx}
\usepackage{mathrsfs}
\usepackage{ulem}
\usepackage{dcolumn}
\usepackage{bm}
\usepackage{tensor}
\usepackage{color}
\usepackage{epsfig}
\usepackage{placeins}
\usepackage{textcomp}
\usepackage{bbold}
\usepackage{float}
\usepackage{verbatim}
\usepackage{times}
\usepackage{siunitx}
\usepackage{chemformula}
\usepackage{physics}
\usepackage{multirow}
\usepackage{makecell}
\usepackage{ragged2e}
\usepackage{subcaption}
\usepackage{gensymb}
\usepackage{makecell,booktabs}
\usepackage{color,colortbl}
\usepackage{hyperref}

\makeatletter
\newcommand{\Rmnum}[1]{\expandafter
\@slowromancap\romannumeral #1@}
\makeatother

\renewcommand{\eqref}[1]{\textup{{\normalfont
Eq.~(\ref{#1}}\normalfont)}}

\DeclareCaptionJustification{justified}{\justifying}
\title{Squeezing-enhanced quantum sensing with \\
quadratic optomechanics}

\author[1,\dag]{Sheng-Dian Zhang}
\author[1,\dag]{Jie Wang}
\author[1,\dag]{Qian Zhang}
\author[1]{Ya-Feng Jiao}
\author[1]{Yun-Lan~Zuo}
\author[2]{{\c{S}}ahin~K.~{\"O}zdemir}
\author[3]{Cheng-Wei Qiu}
\author[4,5]{Franco Nori}
\author[1,*]{Hui Jing}

\affil[1]{Key Laboratory of Low-Dimensional
Quantum Structures and Quantum Control of
Ministry of Education,
Department of Physics and Synergetic
Innovation Center for Quantum Effects
and Applications,
Hunan Normal University,
Changsha 410081, China}
\affil[2]{Department of Engineering Science
and Mechanics, and Materials Research Institute,
Pennsylvania State University, University Park,
State College, Pennsylvania 16802, USA}
\affil[3]{Department of Electrical and
	Computer Engineering, National University of
	Singapore, Singapore 117583, Singapore}
\affil[4]{Theoretical Quantum Physics
	Laboratory, RIKEN Cluster for Pioneering
	Research, Wako-shi, Saitama 351-0198, Japan}
\affil[5]{Physics Department, The University
	of Michigan, Ann Arbor, Michigan 48109-1040, USA}
\affil[$\dag$]{The authors contributed equally to this work}
\affil[*]{jinghui73@foxmail.com}

\begin{abstract}
Cavity optomechanical (COM) sensors, enhanced by quantum squeezing or entanglement, have become powerful tools for measuring ultra-weak forces with high precision and sensitivity. However, these sensors usually rely on linear COM couplings, a fundamental limitation when measurements of the mechanical energy are desired. Very recently,  a giant enhancement of the signal-to-noise ratio was predicted in a quadratic COM system. Here we show that the performance of such a system can be further improved surpassing the standard quantum limit by using quantum squeezed light. Our approach is compatible with  available engineering techniques of advanced COM sensors and provides new opportunities for using COM sensors in tests of fundamental laws of physics and quantum metrology
applications.
\end{abstract}

\setboolean{displaycopyright}{false} 

\begin{document}

\maketitle

\section{Introduction}

The development of quantum-enhanced sensors aimed at the sensitive measurement of time, temperature, pressure, or electromagnetic fields has witnessed considerable progress in recent years~\cite{Clerk2010Introduction,degen2017quantum}, with a broad spectrum of approaches including the use of elementary particles~\cite{pedrozo2020entanglement, mccormick2019quantum,wu2023quantum,chalopin2018quantum}, superconducting circuits~\cite{castellanos2008amplification, backes2021quantum,Xu2022Metrological}, optical systems~\cite{ schnabel2017squeezed,lawrie2019quantum, wang2022dissipation}, and solid-state mechanical devices~\cite{chen2023quantum,gilmore2021quantum}. In particular, cavity optomechanical (COM)~\cite{gavartin2012hybrid, tse2019quantum,yap2020broadband} and electromechanical sensors~\cite{clark2016observation,Wollman2015quantum,pirkkalainen2015squeezing} are remarkably well suited for the measurement of weak forces or very small displacements~\cite{LiOuLeiLiu2021}. Importantly, their standard quantum limit (SQL), which results from the combined effects of backaction noise and photon shot noise, can be broken by the use of optical fields with appropriate quantum correlations (see e.g.~\cite{kampel2017improving,
mason2019continuous}). For example, in an impressive recent experiment, the sub-SQL displacement measurement in a COM system with a macroscopic $\SI{40}{\kilo\gram}$ mirror was achieved by injecting squeezed light in the otherwise empty port of the system~\cite{yu2020quantum}.

COM displacement sensors typically rely on the linear coupling between the displacement of the mechanical element and the electromagnetic field. However, such a coupling is not appropriate for energy or phonon number measurements, which require instead an optomechanical coupling that is quadratic in the mechanical displacement~\cite{bhattacharya2008optomechanical}.  This coupling also allows for applications such as two-phonon cooling~\cite{ Nunnenkamp2010Cooling}, and a variety of quantum non-demolition (QND) measurements~\cite{jayich2008dispersive,miao2009standard, dellantonio2018quantum,hauer2018phonon,ludwig2012enhanced,clerk2010quantum}. Quadratic COM systems (where the cavity detuning is proportional to the square of the mechanical displacement, i.e., $\omega_{\rm cav}(x)\propto x^2$~\cite{thompson2008strong}) have been demonstrated using, e.g., levitated nanospheres~\cite{ bullier2021quadratic}, membrane-in-the-middle cavities~\cite{thompson2008strong,sankey2010strong,flowers2012fiber,lee2015multimode}, photonic crystals~\cite{paraiso2015position, leijssen2017nonlinear}, and atomic gases~\cite{purdy2010tunable}. Also, selective linear or quadratic COM coupling was achieved via homodyne measurements and utilized to create non-Gaussian mechanical states~\cite{vanner2011selective,brawley2016nonlinear}. However one known issue of quadratic coupling is the linear dissipative coupling typically associated with it and there has been significant  interest in exploiting quantum noise interference to cancel the residual linear backaction in the bad-cavity limit, allowing one to make QND measurements of mechanical energy  using a quadratic COM system~\cite{Yanaybackaction2016}. A recent publication proposed a novel geometry that significantly solves this problem and results in a dramatic reduction of backaction noise~\cite{ dumont2022asymmetry}.

\begin{figure}[ht]
\centering
\includegraphics[width=8.4cm]{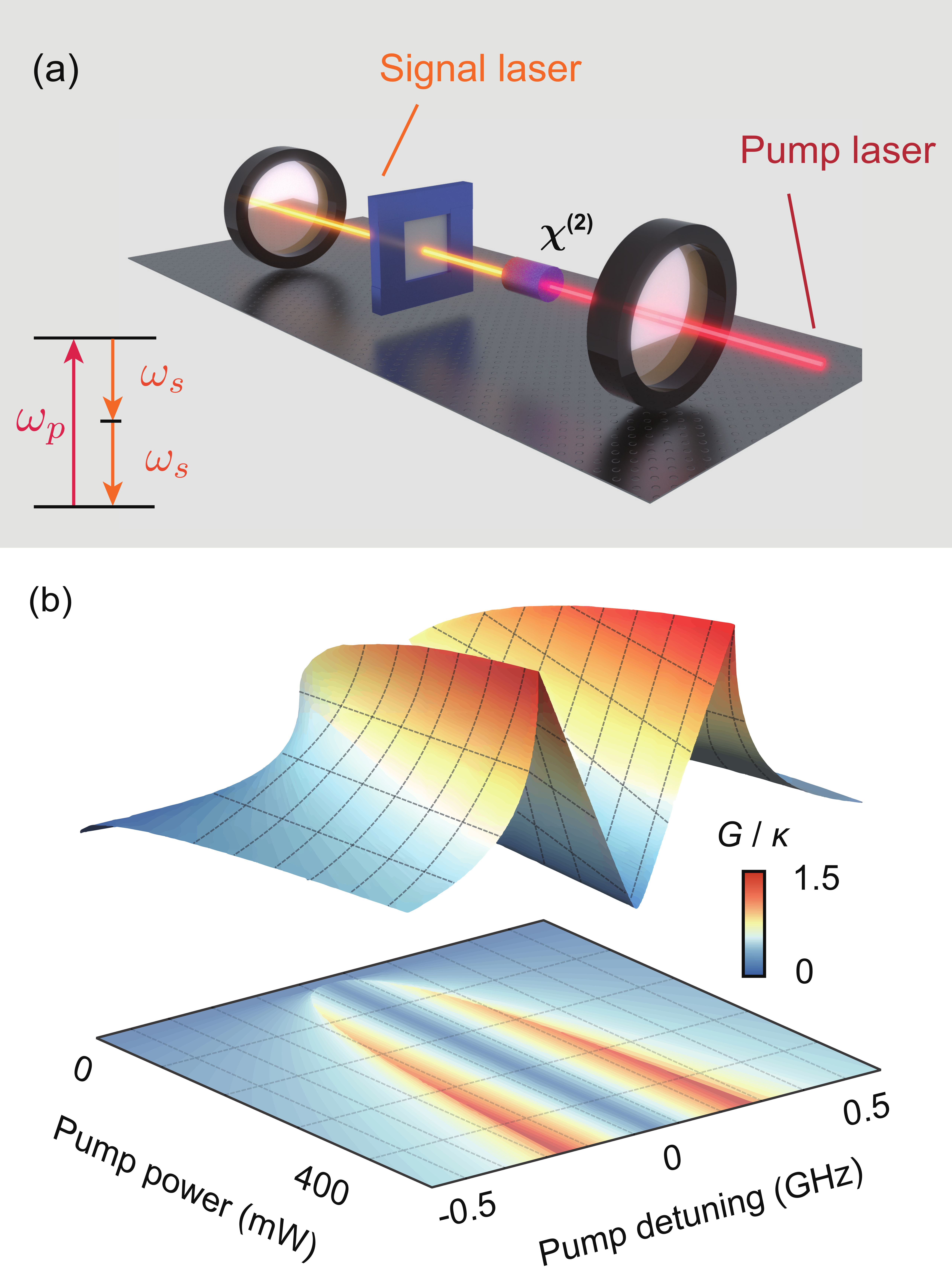}
\caption{Squeezing-enhanced quadratic COM sensing. (a) Schematic representation of the generation of intracavity squeezing within a Fabry-P\'{e}rot cavity. A thin dielectric membrane, at a node or antinode of the standing waves of the cavity, is coupled quadratically with the cavity field, allowing for quantum non-demolition readout of the membrane's phonon numbers~\cite{sankey2010strong}. The nonlinear $\chi^{(2)}$ medium induces intracavity squeezing. (b) The parametric gain $G/\kappa$ versus the pump power and the detuning of the pump mode. Here,  we focus on a red detuning of the pump to generate the required parametric gain~\cite{Peano2015intracavity}. Thus, $G/\kappa$ is decreased at zero-pump detuning, which does not follow the Lorentzian response of the cavity.	}\label{model}
\end{figure}

\begin{figure*}[htb]
\centering
\includegraphics[width=17.5cm]{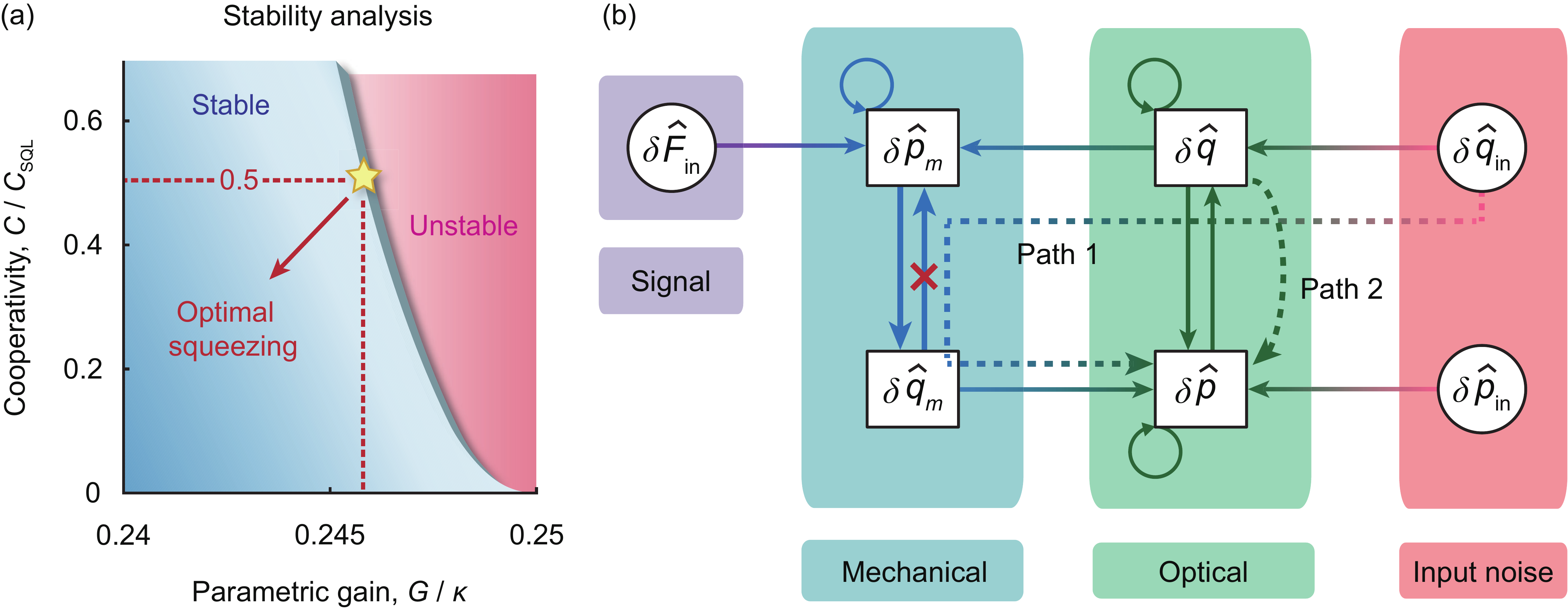}
\caption{(a) System stability versus multi-photon cooperativity and parametric gain (see Supplement 1~\cite{Supplemental} for the details of calculations). The parameters chosen in this paper are confirmed to be within the stable region. The stable and unstable regions can be tuned by altering the parametric gain or the cooperativity, while the optimally sensitive case is close to the border between the stable and unstable regimes. (b) Flow chart representation of~\eqref{noise-eq}, as done in Ref.~\cite{CavesCoherent2010}. $\delta \hat q$ and $\delta \hat p$ are the `position' and `momentum'-like operators of the optical field, respectively, and $\delta \hat q_{\rm in}$ and $\delta \hat p_{\rm in}$ are the associated input noise, respectively. The coefficient from $\delta\hat{{q}}_{m}$ to $\delta\hat{{p}}_{m}$ is zero (indicated by a Red `$\times$'). The original backaction noise in path 1 is eliminated by introducing intracavity squeezing in the antinoise path 2. The path from $\delta\hat{{q}}_{m}$ to $\delta\hat{{p}}_{m}$ is canceled in quadratic COM systems, which results in the enhancement of the destructive interference in the backaction noise. Compared with no squeezing case, the correlation between $\delta\hat{{q}}$ and $\delta\hat{{p}}$ results in the destructive interference, achieving the suppression of the backaction noise. The experimental parameters are chosen as $\Omega_{m}/2\pi= \SI{1}{\mega\hertz}$, $m_{\mathrm{eff}}=\SI{1}{\nano\gram}$, $Q_{m}=\num{5e8}$, $\kappa/2\pi=\SI{3}{\mega\hertz}$. }
\label{motion}
\end{figure*}

In this paper we expand on the study of quadratic optomechanical sensors~\cite{miao2009standard,dumont2022asymmetry,Yanaybackaction2016} and demonstrate theoretically that the inclusion of intracavity optical squeezing~\cite{Exponentially2018Qin,PhysRevLett.129.123602,lu2024quantum} can result in a remarkable improvement in their sensitivity. Our proposed scheme, which is compatible with other available techniques of fabricating and engineering advanced COM sensors, provides a way to further enhance the power of quadratic COM sensors for applications ranging from quantum metrology to tests of fundamental laws of physics.

\section{Squeezed Quadratic Optomechanics}\label{section2}

We consider an ideal membrane-in-the-middle (MIM) Fabry-P\'erot cavity with a thin dielectric membrane located either at a node or antinode of the standing wave mode and coupled quadratically to the field~\cite{thompson2008strong,Meystre2021optics}, allowing for quantum non-demolition readout of the membrane's phonon numbers~\cite{sankey2010strong}. An additional nonlinear $\chi^{(2)}$ medium, coupled quadratically to the cavity field, induces intracavity squeezing, integrated with an intracavity. It is driven by a pump field of frequency $\omega_{p}$ at twice the signal frequency $\omega_{s}$~\cite{Bruch2019opa}, see Fig.~\ref{model}(a). We limit our considerations to the case where the membrane has a low enough reflection that it will not split the cavity into two sub-cavities~\cite{liao2014single, buchmannMacroscopic2012}.

The intracavity second-order nonlinear optical process is described by the Hamiltonian~\cite{guo2016chip}
\begin{align}
\hat H_{\chi^{(2)}} & = \hbar\Delta_c \hat a^{\dagger} \hat a +\hbar\Delta_p \hat a_p^{\dagger} \hat a_p +{\rm i}\hbar \chi^{(2)} \qty( \hat a^{\dagger 2} \hat a_p e^{{\rm i} \theta}
-\hat a^2 \hat a_p^{\dagger} e^{-{\rm i}\theta})\, ,\nonumber\\
\end{align}
where $\hat{a}_{p}$ and $\hat{a}$ are the boson operators of pump and signal modes, of frequencies $\omega_p=2\omega_{s}$; $\Delta_p$ is the detuning between the the pump drive and the nearest cavity mode frequencies; $\Delta_c$ is the detuning between the signal and the nearest cavity mode frequencies, and $\theta$ is the associated phase of $\chi^{(2)}$.
	
We assume that the pump field is strong enough that it can be treated classically, and characterized by a large mean `photon number' $n_p$. Eliminating the associated optomechanical interaction adiabatically  and including the driving ${\cal E}_c$ of the signal mode, we obtain the effective model Hamiltonian at its simplest level~\cite{
thompson2008strong,Peano2015intracavity}:
\begin{align}
\hat{H} & = \hbar\Delta_c\hat a^{\dagger}\hat a +\frac{\hbar}{2}\Omega_m\qty(\hat q_m^{2} +\hat p_m^{2})-\hbar g_{0}\hat a^{\dagger} \hat a\hat q_m^2 \nonumber \\
& \quad +{\rm i}\hbar G\qty(\hat a^{\dagger 2} e^{{\rm i}\theta}-\hat a^2 e^{-{\rm i}\theta}) +{\rm i}\hbar\qty({\cal E}_{c}\hat a^{\dagger}-{\cal E}_{c}^{*}\hat a),\label{H-L eq}
\end{align}
where $\hat{q}_{m}$ and $\hat{p}_{m}$ are the position and momentum operators of mechanical mode at frequency $\Omega_m$; $g_0$ represents single-photon COM coupling strength, which quantifies the interaction between a single phonon and a single photon; $G=\chi^{(2)} \sqrt{n_{p}}$ is the nonlinear gain coefficient, and ${\cal E}_{c}$ is the driving amplitude. We stress that both the quadratic COM coupling and the squeezing-enhanced COM systems were already well-established in experiments. For examples, a high-finesse MIM system was utilized for direct measurements of the membrane's displacement\,\cite{thompson2008strong} and, by tuning the suitable position of the membrane, the quadratic coupling strength can be greatly enhanced for 3 orders of magnitude, indeed reaching a purely quadratic COM system\,\cite{sankey2010strong,Karuza2013}. Such a quadratic COM system was also experimentally demonstrated by levitating a nanosphere in a suitable potential\,\cite{bullier2021quadratic}. We also note that in a recent experiment, by using an intra-cavity parametric amplifier, phase-sensitive manipulations of an input squeezed vacuum was demonstrated\,\cite{zhang2008phase}. Similarly, loss suppressions and thus giant enhancement of sensitivities were also demonstrated in experiments by inserting such optical amplifiers into interferometers\,\cite{vitelli2010,zuo2020}. Indeed, the merits of quantum squeezing in enhancing linear COM sensors have already been confirmed in experiments and the main purpose of our present work is to confirm that such a merit also exists for a quadratic COM system. Hence it is reasonable to expect that even for a hybrid COM system with both linear and quadratic couplings, the positive effects of quantum squeezing will still exist, which we plan to further study in our future work (we note that in a very recent work, the linear coupling was confirmed to be not detrimental for quantum entanglement emerging in such a hybrid COM system \cite{McConnell2024}).
	
Here we use the experimentally feasible parameter values, i.e., the cavity quality factor $Q=\num{1e7}$~\cite{Galinskiy2020Phonon}, the total optical decay rate $\kappa/2\pi=\SI{3}{\mega\hertz}$~\cite{Galinskiy2020Phonon}, including both the decay rate $\kappa_{\mathrm{ex}}$ at the input mirror and the intra-cavity decay rate $\kappa_{\mathrm{0}}$,  with `efficiency'  $\eta_{c}=\kappa_{\mathrm{ex}}/(\kappa_{0}+\kappa_{\mathrm{ex}})$, and the mechanical quality factor $Q_{m}=\num{5e8}$, with the mechanical frequency $\Omega_{m}/2\pi=\SI{1}{\mega\hertz}$~\cite{Galinskiy2020Phonon}, the effective mass $m_{\mathrm{eff}}=\SI{1}{\nano \gram}$~\cite{Galinskiy2020Phonon}, and the associated decay rate $\Gamma_{m}$~\cite{Galinskiy2020Phonon}. We note that a second-order nonlinearity of $\chi^{(2)}/2\pi=\SI{80}{\kilo\hertz}$ was realized~\cite{Bruch2019opa}, confirming the feasibility of $G=0.246\kappa$. Very recently, a new optomechanical experiment using an optical crystal with third-order nonlinearity has demonstrated that with this nonlinearity-assisted system, optical spring effect can be enhanced~\cite{PhysRevLett.132.143602}.
Figure~\ref{model}(b) shows that the nonlinear gain coefficient $G$ increases with the pump laser power and the second-order nonlinearity, indicating the required parametric gain occurs at large pump detunings~\cite{Peano2015intracavity}.
	
Neglecting the higher-order nonlinear terms~\cite{Aspelmeyer2014cavity} in the quantum fluctuations results in coupled linear equations
\begin{align}
\label{noise-eq}
\delta\dot{\hat{a}} & =-\left({\rm i}\Delta + \frac{\kappa}{2}\right)\delta\hat{a} + 2g\delta\hat{q}_{m} +2G e^{{\rm i}\theta}\hat{a}^{\dagger}\nonumber \\
& \,\quad +\sqrt{\eta_{c}\kappa}\delta\hat f_{a,\mathrm{in}} +\sqrt{\left(1-\eta_{c}\right)\kappa} \delta\hat f_{a,0} \, ,\nonumber\\
\delta\dot{\hat{q}}_{m} & =\Omega_{m} \delta\hat p_m,\nonumber \\
\delta\dot{\hat{p}}_{m} &=-\Gamma_m\delta\hat p_m+2g\delta\hat q+\sqrt{2\Gamma_m} \hat{F}_{\mathrm{in}} \, ,
\end{align}
where $g= g_0 \bar q_m |\alpha |$ is the effective optomechanical coupling constant (see Supplement 1 for the detailed classical mean value equations of motion); $\hat f_{a, \mathrm{in}}$  and $\hat f_{a,\mathrm{0}}$ are the noise operators associated with the input cavity mirror and the internal losses, and $\Delta =\Delta_c-g_0\bar q_m^2$ is the effective optical detuning. The flowchart of Fig.~\ref{motion}(b) illustrates the various couplings involved in ~\eqref{noise-eq} . A variable on the right-hand side of an equation of motion is connected to a variable on the left-hand side by arrows, showing that $\delta\dot{\hat p }_m$ is indeed independent of $\delta{\hat{q}}_m$, a consequence of the cancellation of the associated coefficient, i.e., $-\Omega_{m}+2g_{0}n_{c}=0$, where $n_c=\Omega_{m}/(2g_0)$.

Direct measurements of intracavity fields are typically challenging, and one often measures the field that escapes the resonator instead. The relationship between the input field and the outout field is given by the input-output relation $\hat a_{\mathrm{out}}=\sqrt{\eta_c\kappa} \hat a-\hat a_{\mathrm{in}}$ \cite{Aspelmeyer2014cavity}.
As illustrated in Fig.~\ref{motion}(a-b), the parameters used in our work are indeed in the optimally sensitive regime at the border between the stable and unstable regions. Figure \ref{motion}(b) shows that in the quadratic COM system under consideration the flow of signal and noise between $\delta\hat q_m$ and $\delta\hat p_m$ is unidirectional, in contrast to the situation for linear COM systems. This causes the mechanical susceptibility of the quadratic COM sensor to differ from the expression $\Omega_m/(\Omega^2_m-\Omega^2 -{\rm i}\Omega\Gamma_m)$ of those systems~\cite{Aspelmeyer2014cavity}.

\begin{figure}[t]
\centering
\includegraphics[width=9cm]{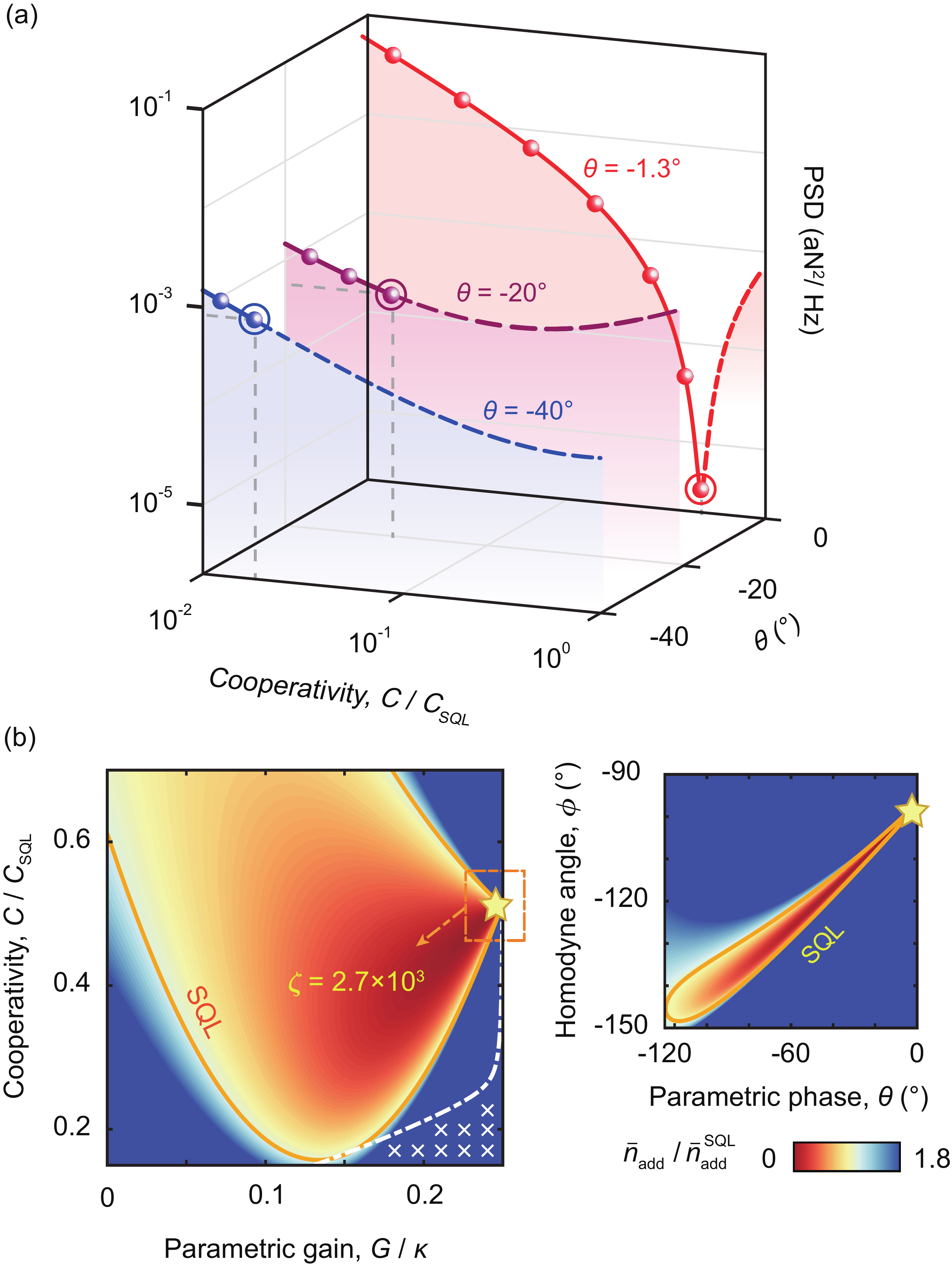}
\caption{Performance of the squeezing-enhanced quadratic COM sensor. (a) Power spectral density (PSD) as a function of  cooperativity and $\theta$. The solid and dashed curves denote the PSD in the stable or unstable region, respectively, and the circles give the minimum PSD in the stable region. The mechanical parameters are $\Omega_{m}/2\pi=\SI{1}{\mega\hertz}$, $m_{\mathrm{eff}}=\SI{1}{\nano\gram}$,  $Q_{m}=\num{5e8}$. (b) Quantum noise below the SQL, with suitable squeezed parameters. The white dashed curve denotes the mechanical response of $\mathcal{R}_{m}\qty[\Omega]=1$. The signal is amplified except for the marked region (`$\times$'). Added noise $\bar{n}_{\mathrm{add}}/ \bar{n}_{\mathrm{add}}^{\mathrm{SQL}}$ as a function of the phase of the local oscillator $\phi$ and the parametric phase $\theta$, and the remaining parameters are $G/\kappa=0.246$, $C/C_{\mathrm{SQL}}=0.5$.}
\label{sql}
\end{figure}

One way to measure the frequency-dependent force noise is  homodyne detection~\cite{Sudhir2017}, whereby
the output signal is mixed at a 50:50 beam splitter with a local oscillator, with a phase $\phi$ between the signal and the reference field.
The photocurrent $\hat{I}_{\phi}$ at the output of the balanced detector is then proportional to a rotated field quadrature
\begin{equation}
\delta\hat{q}_{\mathrm{out}}^\phi [\Omega]=\delta\hat q_{\mathrm{out}} \cos\phi+\delta\hat{p}_{\mathrm{out}}\sin\phi.
\end{equation}
Introducing the correlation functions~\cite{gardiner2004quantum}
\begin{align}
& \left\langle\delta\hat{q}_{u}
\left[\omega\right]\delta\hat{q}_{u}
\left[\Omega\right]\right\rangle
=\left\langle \delta\hat{p}_{u}
\left[\omega\right]\delta\hat{p}_{u}
\left[\Omega\right]\right\rangle
=\frac{1}{2}\delta\left(\omega
+\Omega\right),\nonumber \\
&\left\langle\delta\hat{q}_{u}
\left[\omega\right]\delta\hat{p}_{u}
\left[\Omega\right]\right\rangle
=-\left\langle \delta\hat{p}_{u}
\left[\omega\right]\delta\hat{q}_{u}
\left[\Omega\right]\right\rangle
=\frac{{\rm i}}{2}\delta\left(\omega
+\Omega\right),\nonumber \\
& \left\langle \delta\hat{F}_{\mathrm{th}}
\left[\omega\right]\delta\hat{F}_{\mathrm{th}}
\left[\Omega\right]\right\rangle
=\bar{n}_{m} \delta\left(\omega
+\Omega\right),
\end{align}
where $u=\mathrm{in},~0$ and $\bar{n}_{m}=\qty[\exp(\hbar\Omega_{m}/k_{B}T)-1]^{-1}$ denotes the thermal phonon occupancy. The output amplitude and phase quadrature spectrum can be expressed as~\cite{Sudhir2017}
\begin{align}
\bar{S}_{\mathrm{qq}}^{\mathrm{out}}\qty[\Omega]
& =\frac{1}{2}\left\langle\acomm{\delta\hat{q}_{\mathrm{out}}
\qty[\Omega]}{\delta\hat{q}_{\mathrm{out}}\qty[-\Omega]}
\right\rangle \nonumber\\
& =\frac{1}{2}\mathcal{K}_{-}\qty[\Omega]
+\bar{n}_{m}\abs{\mathcal{N}_{-}\qty[\Omega]}^{2},
\nonumber\\
\bar{S}_{\mathrm{pp}}^{\mathrm{out}}\qty[\Omega]
& =\frac{1}{2}\left\langle\acomm{\delta\hat{p}_{\mathrm{out}}
\qty[\Omega]}{\delta\hat{p}_{\mathrm{out}}\qty[-\Omega]}
\right\rangle \nonumber\\
& =\frac{1}{2}\mathcal{K}_{+}\qty[\Omega]
+\bar{n}_{m}\abs{\mathcal{N}_{+}\qty[\Omega]}^{2},
\end{align}
in the above two equations, we introduced the following definitions
\begin{align} &\mathcal{K}_{-}=\abs{\mathcal{A}_{-}}^{2}
+\abs{\mathcal{B}_{-}}^{2}+\abs{\mathcal{C}_{-}}^{2}
+\abs{\mathcal{D}_{-}}^{2},\nonumber\\
&\mathcal{K}_{+}=\abs{\mathcal{A}_{+}}^{2}
+\abs{\mathcal{B}_{+}}^{2}+\abs{\mathcal{C}_{+}}^{2}+\abs{\mathcal{D}_{+}}^{2}.
\end{align}
The parameters $\mathcal{A}_{\pm}$, $\mathcal{B}_{\pm}$, $\mathcal{C}_{\pm}$, $\mathcal{D}_{\pm}$, and $\mathcal{N}_{\pm}$ can be derived through straightforward algebraic calculations (see Supplement 1~\cite{Supplemental} for their lengthy expressions). $\mathcal{K}_{\pm}$ denotes the contributions of shot noise and backaction noise to the output amplitude or phase quadrature spectrum $\bar{S}_{\mathrm{qq,pp}}^{\mathrm{out}}$, while $\mathcal{N}_{\pm}$ is from the noise imprinted by mechanical motion.
The symmetrized cross-correlation spectrum is then written as
\begin{align}
\bar{S}_{\mathrm{pq}}^{\mathrm{out}}\qty[\Omega]
& =\frac{1}{2}\left\langle\acomm{\delta\hat{q}_{\mathrm{out}}
\qty[\Omega]}{\delta\hat{p}_{\mathrm{out}}\qty[-\Omega]}
\right\rangle \nonumber\\
&=\Re\left \{\frac{1}{2} \mathcal{K}_{\mathrm{co}} \qty[\Omega]+\bar{n}_{m}\mathcal{N}\qty[\Omega] \right\},
\end{align}
with
$\mathcal{K}_{\mathrm{cr}}=\mathcal{B}_{-}\mathcal{A}_{+}^{*}
-\mathcal{A}_{-}\mathcal{B}_{+}^{*}+\mathcal{D}_{-}
\mathcal{C}_{+}^{*}-\mathcal{C}_{-}\mathcal{D}_{+}^{*},$ and
$\mathcal{K}_{\mathrm{si}}=\mathcal{A}_{-}\mathcal{A}_{+}^{*}
+\mathcal{B}_{-}\mathcal{B}_{+}^{*}+\mathcal{C}_{-}
\mathcal{C}_{+}^{*}+\mathcal{D}_{-}\mathcal{D}_{+}^{*}.$
Here $\mathcal{K}_{\mathrm{co}}=\mathcal{K}_{\mathrm{cr}}
+\rm i\mathcal{K}_{\mathrm{si}}$, which contains the squeezed-dependent correlations between shot noise and backaction noise, and $\mathcal{N}=\mathcal{N}_{+}^{*}\mathcal{N}_{-}$~\cite{purdy2013strong}.
The output spectrum thus contains amplitude or phase vacuum noises,
thermal occupations, and quantum correlations~\cite{Sudhir2017},
viz.,
\begin{align}
\bar{S}_{\mathrm{II}}\qty[\Omega]& = \frac{1}{2}\langle
\acomm*{\delta\hat{q}_{\mathrm{out}}^{\phi}\qty[\Omega]}{\delta
\hat{q}_{\mathrm{out}}^{\phi}\qty[-\Omega]}\rangle\nonumber\\
& = \bar{S}_{\mathrm{qq}}^{\mathrm{out}}\cos^{2}\phi
+ \bar{S}_{\mathrm{pp}}^{\mathrm{out}}\sin^{2}\phi
+ \bar{S}_{\mathrm{pq}}^{\mathrm{out}}\sin\qty(2\phi) \nonumber\\
& = \mathcal{R}_{m}\qty[\Omega](\bar{n}_{m}+\bar{n}_{\mathrm{add}}
\qty[\Omega]).
\end{align}
By tuning the squeezed parameters $G$ and $\theta$, the cross term $\mathcal{K}_{\mathrm{co}}$ in $\bar{S}_{\mathrm{pq}}^{\mathrm{out}} $ can become negative, allowing for cancellation of backaction noise and shot noise. The mechanical response of our quadratic COM sensor to the
detected force signal is derived as
\begin{equation}\label{R_m}
\mathcal{R}_{m}= \abs{\mathcal{N}_{-}}^{2}\cos^{2}
\phi+ \abs{\mathcal{N}_{+}}^{2}\sin^{2}\phi
+\Re{\mathcal{N}}\sin\qty(2\phi).
\end{equation}
The value of $\mathcal{R}_{m}$ can be tuned with the squeezing parameters $G$ and $\theta$, leading to effective amplification of the force signal when $\mathcal{R}_{m} > 1$~\cite{Levitan2016mpa}.
The added noise is
\begin{equation}\label{added}
	\bar{n}_{\mathrm{add}}=\frac{\mathcal{K}_{-}
		\cos^{2}\phi+\mathcal{K}_{+}\sin^{2}\phi+\Re{\mathcal{K}_{\mathrm{co}}}\sin\qty(2\phi)}{2\mathcal{R}_{m}}.
\end{equation}
The added noise includes both the shot noise
and the backaction noise, contributing to the total force noise
spectrum for quantifying the sensitivity of the force measurement
\begin{equation}
\bar{S}_{\mathrm{FF}}\qty[\Omega]=2\hbar m_{\mathrm{eff}}
\Gamma_{m}\Omega_{m}(\bar{n}_{m}+\bar{n}_{\mathrm{add}}).
\end{equation}
As detailed in Ref.~\cite{mason2019continuous}, quantum correlations, arranging destructive interference of the imprecision noise and the quantum backaction noise, can be observed in the measured spectrum by detecting rotated quadratures, including amplitude and phase fluctuations, as opposed to standard phase measurements.  Also, the thermal noise, subtracted to reveal quantum noise, can be suppressed by considering a feasible bath temperature  of $\SI{.2}{\kelvin}$ with a cavity placed inside a dilution refrigerator~\cite{fogliano2021ultrasensitive}.

\section{Squeezing-enhanced sensing}\label{section5}

\begin{figure*}[t]
\centering
\includegraphics[width=17.5cm]{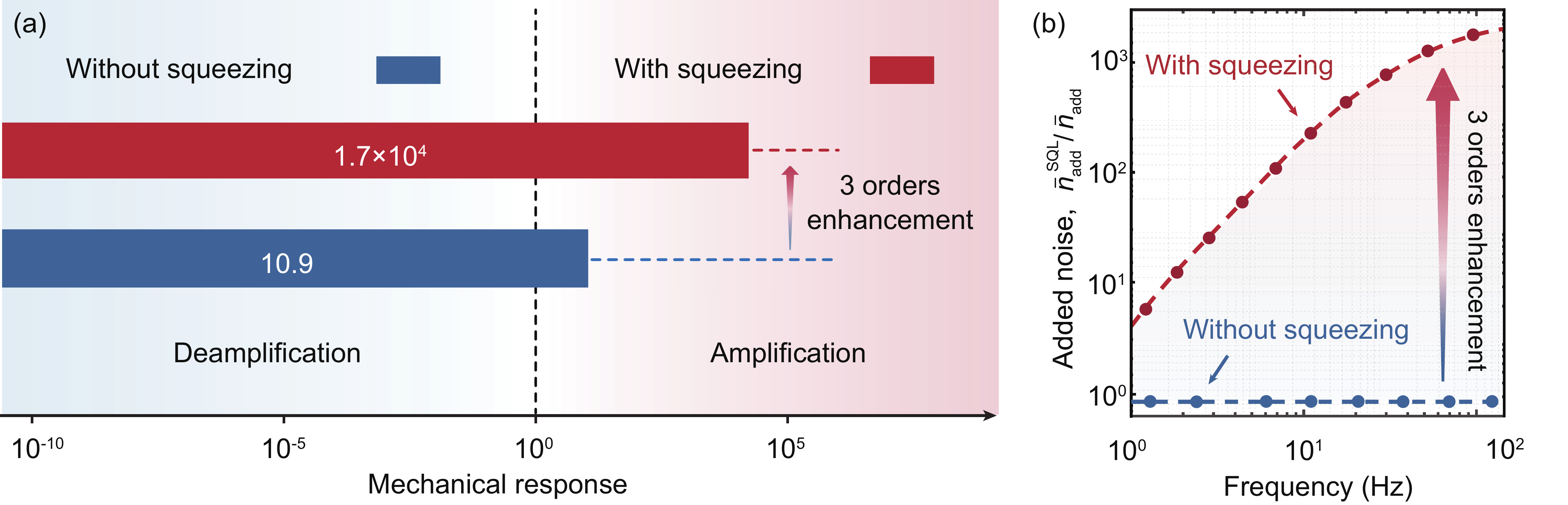}
\caption{(a) A comparison of the optimal mechanical responses for standard and squeezed quadratic COM sensors. The mechanical response describes the amplification ($\mathcal{R}_{m}>1$) or the deamplification ($\mathcal{R}_{m}<1$) of the force signal imprinted on the output quadratures~\cite{Levitan2016mpa}. The optimal mechanical response for the quadratic COM sensor can be further enhanced by introducing intracavity squeezing. (b) For the quadratic COM sensors, the force sensitivity can be significantly enhanced by intracavity squeezing (with the same experimentally accessible parameters)~\cite{Galinskiy2020Phonon}. The remaining parameters are $\Omega_{m}/2\pi=\SI{1}{\mega\hertz}			$, $m_{\mathrm{eff}}=\SI{1}{\nano\gram}$,  $Q_{m}=\num{5e8}$ , $\kappa/2\pi=\SI{3}{\mega\hertz}$, $G/\kappa=0.246$, $C/C_{\mathrm{SQL}}=0.5$, 							$\theta=1.3\degree$, and $\phi=-98\degree$. We produce Fig.~\ref{response}(a-b) according to~\eqref{R_m} and ~\eqref{enhanced}, respectively. }
	\label{response}
\end{figure*}	

Quantum squeezing is known to be capable of increasing the COM sensitivity~\cite{lawrie2019quantum}. In the absence of a $\chi^{(2)}$ medium (or with $\theta = 0$), the force sensitivity is limited by the SQL where in the limit of $\kappa\gg\Omega$ the symmetrized noise spectrum takes the simplified form
\begin{equation}
\bar{n}_{\mathrm{add}}\qty[\Omega]=C+\frac{1}{16\eta_{c}C \Gamma_{m}^{2}\left|\chi_{m}\right|^{2}},
\end{equation}
where the multi-photon cooperativity is defined as
\begin{equation}
C=\frac{4g^{2}}{\kappa\Gamma_{m}}.
\end{equation}
It is minimized to
\begin{equation}
\bar{n}_{\mathrm{add}}^{\mathrm{SQL}}\qty[\Omega] =\frac{1}{2\sqrt{\eta_c} \Gamma_{m}\abs{\chi_m}},
\end{equation}
for
\begin{equation}
C\equiv C_{\mathrm{SQL}}=\frac{1}{4\sqrt{\eta_c} \Gamma_m |\chi_m| } ,
\end{equation}
where $\chi_m=-\Omega_{m}/\qty(\Omega^{2}+{\rm i}\Omega\Gamma_{m})$ is the mechanical susceptibility of the system, which quantifies the response of the oscillator to external forces. So that in the absence of squeezing the minimum output force noise is given by
\begin{equation}
\bar{S}_{\mathrm{FF}}^{\mathrm{SQL}}[\Omega]=2\hbar m_{ \mathrm{eff}}\Gamma_m\Omega_m(\bar n_m+\bar{n}_{\mathrm{add}}^{
\mathrm{SQL}}).
\end{equation}
It is clear from Fig.~\ref{motion}(b) that a direct way to counter the effect of the backaction noise is to introduce another path from $\delta\hat{q}$ to $\delta\hat{p}$ using intracavity squeezing~\cite{CavesCoherent2010}. Then, without standard phase detection, the imprecision and backaction noises can be correlated by tuning the parametric phase of $\chi^{(2)}$ medium. Thus, a decreased parametric phase corresponds to a lower detection sensitivity in the stable region because of the narrower range for the multi-photon cooperativity [Fig.~\ref{sql}(a)].

To simultaneously achieve quantum noise suppression and force signal amplification, the values of the scaled cooperativity ($C / C_{\mathrm{SQL}}$) and the squeezed parameters should be chosen within the stable region [Fig.~\ref{sql}(b)]. For the parameters of our numerical examples it yields the quantum noise that is $3.5$ decibels below the SQL [see Fig.~\ref{sql}(b)]. We note that in a very recent experiment, using a linear COM system assisted by quantum correlations~\cite{yu2020quantum}, a joint quantum uncertainty that is 3 decibels below the SQL was shown after subtracting thermal noises. Here, we define the degree of the squeezing as
\begin{equation}
\begin{aligned}
\sigma=\lg\left(S_{{\rm FF}}/S_{{\rm SQL}}\right).
\end{aligned}
\end{equation}
Quantum-enhanced force measurement can be simply characterized by the enhancement factor due to the squeezing
\begin{equation}\label{enhanced}
\zeta = \frac{\min\qty{\bar{S}_{\mathrm{FF}}\left(G=0,\theta=0
\right)}}{\min\qty{\bar{S}_{\mathrm{FF}}\left(G\neq 0,\theta
\neq 0\right)}}.
\end{equation}
When the thermal noise of the system has been significantly reduced, for instance by utilizing dilution refrigeration or precooling, further enhancement could be further improved by injecting squeezed vacuum into the optical cavity~\cite{squeezing2015Lyu,Exponentially2018Qin}.

Notably, the mechanical susceptibility can transduce force into the displacement of the membrane and quantify the response of the mechanical resonator to the detected force~\cite{Aspelmeyer2014cavity}. In the quadratic COM system, the mechanical response [derived from \eqref{R_m}] to the detected force is significantly enhanced [Fig.~\ref{response}(a)]  due to the larger mechanical susceptibility, enabling a remarkable amplification of the force signal and corresponding to a low quantum noise given by~\eqref{added}.  Hence, the enhanced mechanical response is important to achieve better measurement sensitivity. As shown in Fig.~\ref{response}(a), the optimal mechanical response derived from \eqref{R_m} for the quadratic COM sensor can be further enhanced by introducing intracavity squeezing. Therefore, from the analyses made above, according to \eqref{enhanced}, combined with the additional merit of quantum squeezing, the quadratic COM systems can be more beneficial by incorporating the additional merit of quantum squeezing [Fig.~\ref{response}(b)].

The advantage of the quadratic COM system is mainly manifested in quantum-noise-dominated situations, which becomes marginal with increasing thermal noises. The high sensitivity is predicted close to the boundary between the stable and unstable regimes~\cite{Note1}, as shown in Fig.~\ref{motion}(a). The sensitivity of force measurements is mainly limited by the thermal Langevin force, with the PSD given by~\cite{tsaturyan2017ultracoherent}
\begin{equation}
\bar{S}_{\mathrm{FF}}^{\mathrm{th}}=2m_{\mathrm{eff}}k_{\mathrm{B}}T
\frac{\Omega_{m}}{Q_{m}},
\end{equation}
where $k_{\mathrm{B}}$ is Boltzmann's constant, and $T$ is the bath temperature.
In practice, thermal noise can lower the measurement sensitivity. Nevertheless, we estimate that under realistic conditions, the force sensitivity can still reach $\qty(\SI{10.2}{\atto\newton})^{2}/\si{\hertz}$ even at room temperature (which can be optimized as $\qty(\SI{0.26}{\atto \newton})^{2}/\si{\hertz}$ at cryogenic temperatures), approaching the level of the state-of-the-art sensors with force noises in the range $10$\,--\,$100~\si{\atto\newton.\hertz^{-1/2}}$ at room temperature (or less than $\SI{1}{\atto\newton.\hertz^{-1/2}}$ at cryogenic temperatures)~\cite{David2020microscopy}.  We estimate that by using the state-of-the-art membrane~\cite{ meta2021kippenberg}, the force noise even can be reduced to $\qty(\SI{9.9}{\zepto\newton}) ^{2}/\si{\hertz}$ at the temperature of $\SI{0.2}{\kelvin}$.

For a highly reflective membrane, another practical concern is the backaction arising from the underlying linearity of hybridized
modes~\cite{miao2009standard,Yanaybackaction2016, dumont2022asymmetry}. However, this technical challenge has not prevented the advances in quadratic COM systems~\cite{miao2009standard,Yanaybackaction2016, dumont2022asymmetry}. In fact, linear backaction can be effectively suppressed in practice through structural design or active feedback~\cite{dumont2022asymmetry, brawley2016nonlinear}, or by using highly tunable COM systems such as levitated particles, photonic crystals, electromechanical devices, and cold atoms~\cite{bullier2021quadratic, purdy2010tunable,leijssen2017nonlinear,ma2021non,burgwal2023enhanced}. Indeed, the merits of quantum squeezing in enhancing linear COM sensors have already been confirmed in experiments and the main purpose of our present work is to confirm that such a merit also exists for a quadratic COM system. Hence it is reasonable to expect that even for a hybrid COM system with both linear and quadratic couplings, the positive effects of quantum squeezing will still exist---a specific topic we plan to further calculate and verfiy in our next work.

\section{Conclusion}\label{section6}

In summary, we have shown that the performance of quadratic COM sensors can be significantly enhanced by intracavity squeezing. We find that the mechanical response to weak force signals can be significantly amplified with considerably reduced quantum noise in these systems, promising sub-SQL force measurements with experimentally accessible parameters. We expect that by combining it with other existing techniques of fabricating and operating COM-based sensors, such as those involving feedback control~\cite{gavartin2012hybrid,harris2013minimum} or advanced materials with much higher mechanical $Q$ factors~\cite{meta2021kippenberg,crystal2021kippenberg}, it is possible to further improve its performance in practice. Such an improved COM sensor can be useful for a wide range of applications requiring ultrahigh sensitivity~\cite{Robust2020Zheng,ockeloen2017noiseless,gao2017optical, lecocq2015quantum,safavi2013squeezed,stoev2021highly}. It is our hope that these results will stimulate further efforts toward building and utilizing quantum-squeezing-enhanced sensors, such as those based on levitated spheres~\cite{bullier2021quadratic}, cold atoms~\cite{purdy2010tunable}, dissipative or near-field COM systems~\cite{brawley2016nonlinear}.

\begin{backmatter}
\bmsection{Funding} National Key Research and Development Program of China
(2024YFE0102400); National Natural Science Foundation of China
(12147156); National Natural Science Foundation of China (11774086,
11935006); Multidisciplinary University Research Initiative
(FA9550-21-1-0202); the Foundational Questions Institute Fund (FQXi) (FQXi-IAF19-06);
the Asian Office of Aerospace Research and Development (AOARD)
(FA2386-20-1-4069); the Japan Society for the Promotion of Science (JSPS)
via the Grants-in-Aid for Scientific Research (JP20H00134); the Ministry
of Education, Singapore (A-0005143-01-00); the Science and Technology
Innovation Program of Hunan Province (2021RC2078); Hunan provincial
major sci-tech program (2023ZJ1010); the Science and Technology
Innovation Program of Hunan Province (2020RC4047).

\bmsection{Acknowledgments} We thank Pierre Meystre for helpful discussions and good suggestions.

\bmsection{Disclosures} The authors declare no conflicts of interest.

\bmsection{Data availability} Data underlying the results presented in this paper may be obtained from the authors upon reasonable request.

\bmsection{Supplemental document} See Supplement 1 for supporting content.

\end{backmatter}

\onecolumn

\newpage

\setcounter{equation}{0}
\setcounter{figure}{0}
\setcounter{table}{0}
\setcounter{section}{0}
\renewcommand{\theequation}{S\arabic{equation}}
\renewcommand{\thefigure}{S\arabic{figure}}
\renewcommand{\bibnumfmt}[1]{[S#1]}
\renewcommand{\citenumfont}[1]{S#1}
\renewcommand\thesection{S\arabic{section}}
\renewcommand{\thetable}{S\arabic{table}}

\newcommand{\rme}{\mathrm{e}}
\newcommand{\rmi}{\mathrm{i}}
\newcommand{\rmd}{\mathrm{d}}

\begin{center}
		{\large \bf Supplemental Material for \\
			\vspace{0.5em}
			``Squeezing-Enhanced Quantum Sensing with Quadratic Optomechanics''}
	\end{center}

	\begin{center}
		Sheng-Dian Zhang,$^{1,*}$ Jie Wang,$^{1,*}$
		Qian Zhang,$^{1,*}$ Ya-Feng Jiao,$^{1}$ Yun-Lan Zuo,$^{1}$ \\ \vspace{0.1em}
		{\c{S}}ahin~K.~{\"O}zdemir,$^{2}$
		Cheng-Wei Qiu,$^{3}$ Franco Nori,$^{4,\,5}
		$ and Hui Jing$^{1,\dagger}$
	\end{center}

	\begin{minipage}[]{16cm}
		\small{\it
			\centering $^{1}$Key Laboratory of Low-Dimensional
			Quantum Structures and Quantum Control of
			Ministry of Education, \\
			\centering Department of Physics and Synergetic
			Innovation Center for Quantum Effects \\
			\centering and Applications, Hunan Normal
			University, Changsha 410081, China \\
			\centering $^{2}$Department of Engineering Science
			and Mechanics, and Materials Research Institute, \\
			\centering Pennsylvania State University,
			University Park, State College, Pennsylvania
			16802, USA \\
			\centering $^{3}$Department of Electrical and
			Computer Engineering, \\
			National University of Singapore, Singapore
			117583, Singapore \\
			\centering $^{4}$Theoretical Quantum Physics
			Laboratory, RIKEN Cluster for Pioneering
			Research, Wako-shi, Saitama 351-0198, Japan \\
			\centering $^{5}$Physics Department, The University
			of Michigan, Ann Arbor, Michigan 48109-1040, USA \\
			\centering $^{*}$ These authors contribute equally to this work \\
			\centering $^{\dagger}$ To whom correspondence should
			be addressed; E-mail:  jinghui73@foxmail.com \\
			\rm (Dated: \today) \\
		}
	\end{minipage}
	\vspace{8mm}

	Here, we present more technical details on
	quantum-squeezing-enhanced quadratic optomechanical sensing, including:
	(1) detailed derivations of the linearized
	Hamiltonian;
    (2) the output noise spectrum;
	(3) degenerate optical parametric oscillations;
	(4) discussions on
	stability conditions;
	(5) signal-to-noise ratio and the
	optimal variance of the rotated field quadrature;
	(6) extended applications to the state-of-the-art
	quantum sensors.
	
	\section{Derivation of the linearized
		Hamiltonian}
	
	In our cavity optomechanical (COM) system,
	the pump laser with frequency $\omega_{p}$
	has twice the frequency of the signal laser ($\omega_{s}$).
	Each laser tone (pump and signal)
	is quasi-resonant with a particular optical normal mode of the a Fabry-P\'{e}rot cavity, thus we
	refer to these optical modes as pump and signal mode, respectively~\cite{
		SI-Peano2015intracavity}.
	The flexible dielectric membrane is placed
	at a location of $\left.q_{0}=j\lambda_{p}
	\middle/4=k\lambda_{s}\middle/4\right.$
	($j$, $k$ integers)~\cite{
		SI-bhattacharya2008optomechanical}, i.e.,
	the common node (or antinode) of the
	intracavity standing waves~\cite{
		SI-sankey2010strong},
	where $\lambda_{p}$ and $\lambda_{s}$
	are the resonant wavelengths for
	the pump and signal modes, respectively.
	We then form a realistic description
	incorporating intrinsic losses and the
	coupling of the mechanical resonator to
	the optical modes, which yields
	the total Hamiltonian in a rotating
	frame~\cite{SI-Liao2014single,
		SI-qin2021generating}:
	\begin{align}
		\hat{H} & = \hbar\Delta_{c}
		\hat{a}^{\dagger}
		\hat{a}+\hbar\Delta_{p}
		\hat{a}_{p}^{\dagger}
		\hat{a}_{p}+\frac{\hbar}{2}\Omega_{m}\qty(
		\hat{q}_{m}^{2}+\hat{p}_{m}^{2})
		-\hbar\hat{q}_{m}^{2}
		\qty(g_{0}\hat{a}^{\dagger}\hat{a}
		+g_{p}\hat{a}_{p}^{\dagger}
		\hat{a}_{p}) \nonumber\\
		& \quad +\rmi\hbar\chi^{(2)}\qty(
		\hat{a}^{\dagger 2}
		\hat{a}_{p}\rme^{\rmi\theta}
		-\hat{a}^{2}\hat{a}_{p}
		^{\dagger}\rme^{-\rmi\theta}
		)+\rmi\hbar\qty({\cal E}_{c}
		\hat{a}^{\dagger}
		+{\cal E}_{p}\hat{a}_{p}^{\dagger}
		-\mathrm{H.C.}),
	\end{align}
	where we  wrote in a frame where the pump and signal modes phase space rotate
	at frequency $\omega_{p}$ and  $\omega_{s}$, respectively, and
	the driving amplitudes are
	$\abs{{\cal E}_{c}}=\sqrt{
		\kappa\eta_{c}P_{c}/
		(\hbar\omega_{s})}$,
	$\abs{{\cal E}_{p}}=\sqrt{
		\kappa_{p}\eta_{p}P_{p}/
		(\hbar\omega_{p})}$.
	The detunings of the optical modes are
	$\Delta_{c}=\Omega_{c}-\omega_{s}$,
	$\Delta_{p}=\Omega_{p}-\omega_{p}$,
	with $g_{0}$ and $g_{p}$ the
	COM coupling strength
	of the signal and pump modes, respectively.
	
	Thus, the equations of motion can be given by
	\begin{align}
		\dot{\hat{a}}&=-\left(i\Delta_{c}+\frac{\kappa}{2}\right)\hat{a}+ig_{0}\hat{a}\hat{q}_{m}^{2}+2\chi^{(2)}\hat{a}^{\dagger}\hat{a}_{p}e^{i\theta}+{\cal E}_{c}, \nonumber\\
		\dot{\hat{a}}_{p}&=-\left(i\Delta_{p}+\frac{\kappa_p}{2}\right)\hat{a}_{p}+ig_{p}\hat{a}_{p}\hat{q}_{m}^{2}-\chi^{(2)}\hat{a}^{2}e^{-i\theta}+{\cal E}_{p},\nonumber\\
		\dot{\hat{q}}_{m}&=\Omega_{m}\hat{p}_{m},\nonumber\\
		\dot{\hat{p}}_{m}&=-\Omega_{m}\hat{q}_{m}-\Gamma_{m}\hat{p}_{m}+2\hat{q}_{m}\left(g_{0}\hat{a}^{\dagger}\hat{a}+g_{p}\hat{a}_{p}^{\dagger}\hat{a}_{p}\right).
	\end{align}
	To proceed, we derive the classical equations for the steady-state values  under the condition of strong optical driving
	\begin{align}
		-\left(i\Delta+\frac{\kappa}{2}\right)\alpha+2\chi^{(2)} e^{i\theta}\alpha\alpha_{p}+\left|{\cal E}_{c}\right|e^{i\Phi}&=0,\nonumber\\
		-\left(i\Delta^{\prime}+\frac{\kappa+p}{2}\right)\alpha_{p}-\chi^{(2)} e^{-i\theta}\alpha^{2}+\left|{\cal E}_{p}\right|e^{i\varphi}&=0,\nonumber\\
		-\Omega_{m}+2g_{0}\alpha^{*}\alpha+2g_{p}\alpha_{p}^{*}\alpha_{p}&=0,
	\end{align}
	where $\Phi$ ($\varphi$) is the phase of the pump
	(signal) laser. Herein, we choose $\frac{\kappa}{2}=\frac{\kappa_p}{2}=\frac{\kappa}{2}$, $g_{0}=g_{0}=\frac{g_{p}}{4}$,
	then the steady-state solutions are
	\begin{align}
		&\begin{aligned}
			2 g_0 \alpha_c^* \alpha_c & =\Omega_m-2 g_p \alpha_p^* \alpha_p, \nonumber\\
			\alpha_c^* \alpha_c & =\frac{\Omega_m-2 g_p \alpha_p^* \alpha_p}{2 g_0}=\frac{\Omega_m-8 g_0 \alpha_p^* \alpha_p}{2 g_0}=\frac{\Omega_m}{2 g_0}-4 \alpha_p^* \alpha_p,
		\end{aligned}\\
		&\begin{aligned}
			-\left(i \Delta+\frac{\kappa}{2}\right) \alpha_c+2 \chi^{(2)} e^{i \theta} \alpha_c \alpha_p+\left|{\cal E}_c\right| e^{i \Phi} & =0,\\
			-\left(i \Delta^{\prime}+\frac{\kappa}{2}\right) \alpha_p-\chi^{(2)} e^{-i \theta}\left(\frac{\Omega_m}{2 g_0}-4\left|\alpha_p\right|^2\right)+\left|{\cal E}_p\right| e^{i \Psi} & =0 \\
			-\sqrt{\Delta^{\prime 2}+\frac{\kappa^2}{4}} \alpha_p-\chi^{(2)}\left(\frac{\Omega_m}{2 g_0}-4 \alpha_p^2\right)+\left|{\cal E}_p\right| & =0 , \\
			4 \chi^{(2)} \alpha_p^2-\sqrt{\Delta^{\prime 2}+\frac{\kappa^2}{4}} \alpha_p+\left|{\cal E}_p\right|-\frac{\chi^{(2)} \Omega_m}{2 g_0} & =0.
		\end{aligned}
	\end{align}
	Thus,
	\begin{align}
		\alpha_{p}&=\frac{\sqrt{\Delta^{\prime2}+\frac{\kappa^{2}}{4}}-\sqrt{\Delta^{\prime2}+\frac{\kappa^{2}}{4}-16\chi^{(2)}\left(\left|{\cal E}_{p}\right|-\frac{\chi^{(2)}\Omega_{m}}{2g_{0}}\right)}}{8\chi^{(2)}}, \nonumber\\
		&=\frac{\sqrt{\Delta^{\prime2}+\frac{\kappa^{2}}{4}}}{8\chi^{(2)}}-\frac{\sqrt{\Delta^{\prime2}+\frac{\kappa^{2}}{4}-16\chi^{(2)}\left(\left|{\cal E}_{p}\right|-\frac{\chi^{(2)}\Omega_{m}}{2g_{0}}\right)}}{8\chi^{(2)}}, \nonumber\\
		&=\frac{1}{8\chi^{(2)}}\sqrt{\frac{\kappa^{2}+4\Delta^{\prime2}}{4}}-\frac{1}{8\chi^{(2)}}\sqrt{\frac{\kappa^{2}+4\Delta^{\prime2}}{4}-16\chi^{(2)}\left(\left|{\cal E}_{p}\right|-\frac{\chi^{(2)}\Omega_{m}}{2g_{0}}\right)}, \nonumber\\
		&=\frac{1}{16\chi^{(2)}}\sqrt{\kappa^{2}+4\Delta^{\prime2}}-\sqrt{\frac{\kappa^{2}+4\Delta^{\prime2}}{4\times64\chi^{(2)2}}-\frac{16\chi^{(2)}}{64\chi^{(2)2}}\left(\left|{\cal E}_{p}\right|-\frac{\chi^{(2)}\Omega_{m}}{2g_{0}}\right)}, \nonumber\\
		&=\frac{1}{16\chi^{(2)}}\sqrt{\kappa^{2}+4\Delta^{\prime2}}-\sqrt{\frac{\kappa^{2}+4\Delta^{\prime2}}{16\times16\chi^{(2)2}}-\frac{1}{4\chi^{(2)}}\left(\left|{\cal E}_{p}\right|-\frac{\chi^{(2)}\Omega_{m}}{2g_{0}}\right)}.
	\end{align}

The external force is described as
	$\hat{F}_{\mathrm{in}}
	=\hat{F}_{\mathrm{th}}+\hat{F}_{\mathrm{sig}}$,
	where $\hat{F}_{\mathrm{th}}
	=\hat{\mathcal{F}}_{\mathrm{th}}
	\big/\sqrt{2\hbar m_{\mathrm{eff}}
		\Gamma_{m}\Omega_{m}}$
	and $\hat{F}_{\mathrm{sig}}
	=\hat{\mathcal{F}}_{\mathrm{sig}}
	\big/\sqrt{2\hbar m_{\mathrm{eff}}
		\Gamma_{m}\Omega_{m}}$
	are the scaled thermal force and the detected
	force signal with dimension \si{\hertz^{\left.1
			\middle/2\right.}}, respectively.
	The variables $\delta
	\hat{a}_{\mathrm{in}}$ and $\delta\hat{a}_{_{0}}$
	represent the fluctuations at the coupling
	port and the port modelling internal losses,
	respectively.
	The single-photon coupling rate is denoted by
	$g_{0}=g_{\mathrm{om}}q_{\mathrm{zp}}^{2}$,
	and the quadratic coupling strength is written as
	$g_{\mathrm{om}}=8\pi^{2}c\sqrt{R/\qty(1-R)}
	/\qty(\lambda_{s}^{2}L)$~\cite{SI-bhattacharya2008optomechanical},
	which can reach $\SI{1.54}{\tera\hertz\per
		\nano\metre^2}$ in experiment~\cite{
		SI-paraiso2015position}.
	Here we define the dimensionless mechanical
	quadratures as $\left.\hat{q}_{m}
	=\hat{q}\middle/q_{\mathrm{zp}}\right.$ and
	$\left.\hat{p}_{m}=\hat{p}\middle/
	p_{\mathrm{zp}}\right.$, where $q_{\mathrm{zp}}
	=\sqrt{\hbar/(m_{\mathrm{eff}}
		\Omega_{m})}$, $p_{\mathrm{zp}}=
	\sqrt{\hbar m_{\mathrm{eff}}\Omega_{m}}$ are the
	standard deviations of the zero-point motion and
	momentum, respectively.  Besides, the signal mode is
	characterized by a total loss rate $\kappa=
	\kappa_{0}+\kappa_{\mathrm{ex}}$ with the efficiency defines as $\eta_{c}=\kappa_{\mathrm{ex}}/
	(\kappa_{0}+\kappa_{\mathrm{ex}})$ describing the contribution of the input coupling loss rate to the total cavity loss rate.
	We note that in a very recent experiment~\cite{
		SI-Bruch2019opa}, the second-order nonlinearity
	was demonstrated with a value of
	$\chi^{(2)}/2\pi=\SI{80}{\kilo\hertz}$.
	Thus, it is possible to generate a nonlinear gain coefficient
	where $\alpha_{p}$ denotes the amplitude of
	the pump mode.
	
	For simplicity, we choose $\alpha_{p}\in\mathbb{R}>0$ and
	take intracavity field as the phase reference,
	i.e., $\alpha\in\mathbb{R}>0$~\cite{
		SI-Peano2015intracavity}. The solutions of the
	steady-state values can thus be expressed as
	\begin{align}\label{steady-eqs}
		\cos\theta &=\frac{1}{4G}\left(\kappa
		-\left|\frac{2{\cal E}_{c}}
		{\sqrt{n_{c}}}\right|\cos\Phi\right),\nonumber\\
		\bar{q}_{m}^{2} &=\frac{1}{g_{0}}\left(\Delta_{c}
		-\left|\frac{{\cal E}_{c}}
		{\sqrt{n_{c}}}\right|\sin\Phi-2G\sin\theta\right),
	\end{align}
	where $n_{c}=\Omega_{m}/\qty(2g_{0})$, and he parametric  phase depends on the phase
	$\Phi$ of the signal laser. Then, the displacement
	of the oscillator is directly proportional to the
	input force signal:
	\begin{equation}
		\delta\hat{q}\qty[\Omega]=\chi_{\mathrm{eff}}
		^{(2)}\qty[\Omega]\delta\hat{\mathcal{F}}
		_{\mathrm{sig}}\qty[\Omega],
	\end{equation}
	where $\chi_{\mathrm{eff}}^{(2)}\qty[\Omega]$
	denotes the effective mechanical susceptibility:
	\begin{align}
		\chi_{\mathrm{eff}}^{(2)}\qty[\Omega]
		&= \frac{1}{m_{\mathrm{eff}}\Omega_{m}
			\qty(\chi_{m}^{-1}+\Sigma)}, \nonumber\\
		\Sigma\qty[\Omega]
		&= \frac{16g^{2}\qty(\Omega-
			\Delta-2G\sin\theta)}{\kappa^{2}-16G^{2}+4\qty(
			\Omega-\Delta)^{2}}.
	\end{align}
	
	For the resonance case without intracavity
	squeezing, the additional term $\Sigma\qty[\Omega]$
	is negligible. The effective susceptibility
	can be written at the simplest level as
	\begin{equation}
		\chi_{\mathrm{eff}}^{(2)}\qty[\Omega]=-\frac{1}
		{m_{\mathrm{eff}}\qty(\Omega^{2}+\rmi\Omega
			\Gamma_{m})}.
	\end{equation}

	Here, we consider the effect of the fluctuations of
	the pump mode. For a strong pump field, this mode can be
	eliminated adiabatically, which yields the shifts of
	the cavity linewidth, the COM coupling rate, and the
	mechanical resonance frequency:
	\begin{equation}\label{SH-flu}
		\kappa_{s}^{\mathrm{eff}} =
		\kappa_{s}+\frac{16\nu^{2}n_{s}}
		{\kappa_{p}+2\rmi\Delta^{\prime}},
		\quad \Omega_{m}^{\mathrm{eff}} =
		-\frac{16G_{p}^{2}\Delta^{\prime}}{
			\kappa_{p}^{2}+4\Delta^{\prime 2}},
		\quad G_{s}^{\pm} = G_{s}\pm\frac{
			4\nu\alpha_{s}G_{p}\rme^{\pm\rmi\theta}}
		{\kappa_{p}+2\rmi\Delta^{\prime}},
	\end{equation}
	where $\Delta=\Delta_{c}-g_{s}\bar{q}_{m}^{2}$,
	$\Delta^{\prime}=\Delta_{p}-g_{p}\bar{q}_{m}^{2}$,
	$G_{s}=g_{s}\bar{q}_{m}\sqrt{2n_{s}}$, and
	$G_{p}=g_{p}\bar{q}_{m}\sqrt{2n_{p}}$.
	Thus, the optical losses are slightly
	modified by the pump mode due to the photon
	up-conversion~\cite{SI-Peano2015intracavity}.
	The additional COM coupling
	and the mechanical eigenfrequency
	indicate the contributions of the
	photon-phonon coupling for the pump
	mode~\cite{SI-Peano2015intracavity}.
	Then, the fluctuations of the pump mode
	can be neglected under a large detuning
	and a small second-order nonlinearity~\cite{
		SI-Peano2015intracavity}.
	
	\section{The output quadratures}
	
	After introducing phenomenologically the various dissipation mechanisms and associated input noise, the Hamiltonian yields readily the quantum Langevin equations
	\begin{align}
		\dot{\hat{a}} & =-\left({\rm i}\Delta_{c} + \frac{\kappa}{2}\right)\hat{a} + {\rm i} g_{0}\hat{a}\hat{q}_{m}^{2} +2G e^{{\rm i}\theta}\hat{a}^{\dagger}\nonumber \\
		&+{\cal E}_{c}+\sqrt{\eta_{c}\kappa}\hat f_{a,\mathrm{in}} +\sqrt{\left(1-\eta_{c}\right)\kappa} \hat f_{a,0} \, ,\nonumber\\
		\dot{\hat{q}}_{m} & =\Omega_{m}\hat{p}_{m}\, , \nonumber \\
		\dot{\hat{p}}_{m} & =-\Omega_{m}\hat{q}_{m} -\Gamma_{m}\hat{p}_{m}+2g_{0}\hat{q}_{m} \hat{a}^{\dagger}\hat{a}+\sqrt{2\Gamma_{m}}\hat{F}_{\mathrm{in}}\, ,\label{QLEs}
	\end{align}
	where $\hat f_{a, \mathrm{in}}$  and $\hat f_{a,\mathrm{0}}$ are the noise operators associated with the input cavity mirror and internal losses, respectively. The noise forces acting on the mechanical membrane are
	\begin{equation}
		\hat{F}_{\mathrm{in}}=\hat{F}_{\mathrm{th}}+\hat{F}_{\mathrm{sig}} \, ,
	\end{equation}
	where $\hat{F}_{\mathrm{th}}$ and $\hat{F}_{\mathrm{sig}}$ are the scaled thermal force and the force signal to be detected, respectively, with dimension \si{\hertz^{\left.1\middle/2\right.}}, respectively. All noise operators have zero mean values
	\begin{equation}
		\begin{aligned}
			\langle\hat f_{a, \mathrm{in}}\rangle = \langle \hat f_{a,\mathrm{0}} \rangle = \langle \hat F_{\mathrm{th}}\rangle = \langle \hat F_{\mathrm{sig}}\rangle = 0.
		\end{aligned}
	\end{equation}
	
	Because of the nonlinear COM interaction, Eqs.~(\ref{QLEs}) do not form a closed set of operator equations. We proceed by considering the situation of a strong driving, and expand each operator as the sum of its classical mean value and a small quantum fluctuation, i.e. $\hat{a}=\alpha+\delta\hat a$,
	$\hat{q}_{m}=\bar{q}_{m}+\delta\hat{q}_{m}$, and $\hat{p}_{m}=\bar{p}_{m}+\delta\hat{p}_{m}$, with $\langle \delta \hat a \rangle = \langle \delta \hat q_m\rangle = \langle \delta \hat p_m\rangle = 0$ .  This yields the classical mean value equations of motion
	\begin{align}
		\dot{\alpha} & =-\left({\rm i}\Delta+\frac{\kappa}{2}\right) \left|\alpha\right|+2G e^{{\rm i}\theta} \left|\alpha\right|+\left|\varepsilon_{c}\right| e^{\Phi},\nonumber \\
		\dot {\bar p}_m & =-\Omega_{m}\bar q_m-\Gamma_{m}\bar p_m +2g_{0}\bar q_m |\alpha|^2\, ,\nonumber \\
		\dot{\bar q}_m & =\bar p_m,
	\end{align}
	where the effective optical detuning is $\Delta =\Delta_c-g_0\bar q_m^2$ and $\Phi$ describes the phase of the driving field. Here we take intracavity field as the phase reference, i.e., $\alpha\in\mathbb{R}>0$, in which case  the steady-state mean values become: $|\alpha|^2 =\Omega_{m}/2g_0$, $\bar p_m =0$, and
	\begin{align}
		\label{steady-eqs}
		|\bar q_m|^2  =\frac{1}{g_0}\left(\Delta_c -\left |\frac{{\cal E}_c} {\alpha}\right| \sin\Phi-2G\sin\theta\right).
	\end{align}

	We now introduce the `position' and `momentum'-like operators of the optical field,
	\begin{align}
		\hat q&=\frac{1}{\sqrt{2}}\left( \hat a^{\dagger}+ \hat a\right),\nonumber \\
		\hat p &=\frac{{\rm i}}{\sqrt{2}}\left(\hat a^{\dagger}-\hat a\right),
	\end{align}
	and the associated optical noise operators
	\begin{align}
		\hat f_{q,\mathrm{in}}&=\frac{1}{\sqrt 2}\left( \hat f^\dagger_{a,\mathrm{in}}+ \hat f_{a,\mathrm{in}} \right), \hat f_{p,\mathrm{in}} =\frac{{\rm i}}{\sqrt{2}}\left( \hat f^\dagger_{a,\mathrm{in}}- \hat f_{a,\mathrm{in}} \right );\nonumber \\
		\hat f_{q,0}&=\frac{1}{\sqrt 2}\left( \hat f^\dagger_{a,0}+ \hat f_{a,0} \right),
		\hat f_{p,0}=\frac{{\rm i}}{\sqrt{2}}\left( \hat f^\dagger_{a,0}- \hat f_{a,0} \right ) .
	\end{align}
	
In the Fourier domain expressions for the output quadratures:
		\begin{align}
			\label{full}
			\hat q_{\mathrm{out}} \left[\Omega\right]&=(\kappa/2-2G\cos\theta-i\Omega)^{-1}\left[\sqrt{\eta_{c}\kappa}\left(\Delta_{c}-2G\sin\theta-g_{0}\hat{q}_{m}^{2}\right)\hat p+\left(\eta_c \kappa-1\right ) \hat f_{q,\mathrm{in}} +\kappa\sqrt{(1-\eta_c)\eta_c} \hat f_{q,0} \right ],\nonumber\\
			\hat{p}_{\mathrm{out}} \left[\Omega\right]&=(\kappa/2+2G\cos\theta-i\Omega)^{-1}\left[-\sqrt{\eta_{c}\kappa}\left(\Delta_{c}-2G\sin\theta-g_{0}\hat{q}_{m}^{2}\right)\hat q +\left(\eta_{c}\kappa-1\right )\hat f_{p,\mathrm{in}}+\kappa\sqrt{(1-\eta_{c})\eta_{c}} \hat f_{p,0} \right].
		\end{align}
		The essential step in quantum sensing is to observe the output fluctuations of physical quantities to be measured in the Fourier domain, i.e.,
		\begin{align}\label{matrix-coefficients}
			\begin{pmatrix}
				\delta \hat{q}_{\mathrm{out}}
				\\
				\delta \hat{p}_{\mathrm{out}}
			\end{pmatrix}=\begin{pmatrix}\mathcal{A}_{-}
				& \mathcal{B}_{-}
				& \mathcal{C}_{-}
				& \mathcal{D}_{-}
				& \mathcal{N}_{-}\\
				-\mathcal{B}_{+}
				& \mathcal{A}_{+}
				& -\mathcal{D}_{+}
				& \mathcal{C}_{+}
				& \mathcal{N}_{+}
			\end{pmatrix}\begin{pmatrix}
				\hat f_{q, \mathrm{in}}
				&\hat f_{p,\mathrm{in}} &
				\hat f_{q, 0} &
				\hat f_{p,0} &
				\hat{F}_{\mathrm{in}}
			\end{pmatrix}^{\mathrm{T}}.
		\end{align}
		where	
		\begin{align}
			\mathcal{A}_{\pm}\qty[\Omega]&=\rho\kappa(\kappa/2\pm 2G\cos\theta-i\Omega)^{-1}\eta_{c}-1, \nonumber\\
			\mathcal{B}_{+}\qty[\Omega]&=\rho\kappa(\Delta-2G\sin\theta-4g^{2}\chi_{\mathrm{m}})(\kappa/2-2G\cos\theta-i\Omega)^{-1}(\kappa/2+2G\cos\theta-i\Omega)^{-1}\eta_{c}\nonumber\\
			\mathcal{B}_{-}\qty[\Omega]&=\rho\kappa(\Delta+2G\sin\theta)(\kappa/2-2G\cos\theta-i\Omega)^{-1}(\kappa/2+2G\cos\theta-i\Omega)^{-1}\eta_{c},\nonumber\\
			\mathcal{C}_{\pm}\qty[\Omega]&=\rho\kappa(\kappa/2 \pm 2G\cos\theta-i\Omega)^{-1}\sqrt{(1-\eta_{c})\eta_{c}},\nonumber\\
			\mathcal{D}_{+}\qty[\Omega]&=\rho\kappa(\Delta-2G\sin\theta-4g^{2}\chi_{\mathrm{m}})(\kappa/2-2G\cos\theta-i\Omega)^{-1}(\kappa/2+2G\cos\theta-i\Omega)^{-1}\sqrt{(1-\eta_{c})\eta_{c}},\nonumber\\
			\mathcal{D}_{-}\qty[\Omega]&=\rho\kappa(\Delta+2G\sin\theta)(\kappa/2-2G\cos\theta-i\Omega)^{-1}(\kappa/2+2G\cos\theta-i\Omega)^{-1}\sqrt{(1-\eta_{c})\eta_{c}},\nonumber\\
			\mathcal{N}_{+}\qty[\Omega]
			& =2g\rho(\kappa/2+2G\cos\theta-i\Omega)^{-1}\chi_{\mathrm{m}}\sqrt{2\kappa\eta_{c}\Gamma_{\mathrm{m}}},\nonumber\\
			\mathcal{N}_{-}\qty[\Omega]
			& =2g\rho(\Delta+2G\sin\theta)(\kappa/2+2G\cos\theta-i\Omega)^{-1}(\kappa/2-2G\cos\theta-i\Omega)^{-1}\chi_{\mathrm{m}}\sqrt{2\kappa\eta_{c}\Gamma_{\mathrm{m}}}\nonumber\\
			\rho&=\left[1+(\kappa/2-2G\cos\theta-i\Omega)^{-1}(\kappa/2+2G\cos\theta-i\Omega)^{-1}( \Delta+2G\sin\theta)\left (\Delta-2G\sin\theta-4g^{2}\chi_{\mathrm{m}}\right )\right]^{-1},
		\end{align}
and $\chi_m=-\Omega_{m}/\qty(\Omega^{2}+{\rm i}\Omega\Gamma_{m})$ is the mechanical susceptibility of the system, which quantifies the response of the oscillator to external forces.  For the case without intracavity squeezing ($G=0$, $\theta=0$), the above coefficients related to the quadratic coupling are
		\begin{align}
		\mathcal{A}^{'}_{\pm}\qty[\Omega]&=\rho\kappa(\kappa/2-i\Omega)^{-1}\eta_{c}-1, \nonumber\\
		\mathcal{B}^{'}_{+}\qty[\Omega]&=\rho\kappa(\Delta-4g^{2}\chi_{\mathrm{m}})(\kappa/2-i\Omega)^{-2}\eta_{c}\nonumber\\
		\mathcal{B}^{'}_{-}\qty[\Omega]&=\rho\kappa\Delta(\kappa/2-i\Omega)^{-2}\eta_{c},\nonumber\\
		\mathcal{C}^{'}_{\pm}\qty[\Omega]&=\rho\kappa(\kappa/2 -i\Omega)^{-1}\sqrt{(1-\eta_{c})\eta_{c}},\nonumber\\
		\mathcal{D}^{'}_{+}\qty[\Omega]&=\rho\kappa(\Delta-4g^{2}\chi_{\mathrm{m}})(\kappa/2-i\Omega)^{-2}\sqrt{(1-\eta_{c})\eta_{c}},\nonumber\\
		\mathcal{D}^{'}_{-}\qty[\Omega]&=\rho\kappa\Delta(\kappa/2-i\Omega)^{-2}\sqrt{(1-\eta_{c})\eta_{c}},\nonumber\\
		\mathcal{N}^{'}_{+}\qty[\Omega]
		& =2g\rho(\kappa/2-i\Omega)^{-1}\chi_{\mathrm{m}}\sqrt{2\kappa\eta_{c}\Gamma_{\mathrm{m}}},\nonumber\\
		\mathcal{N}^{'}_{-}\qty[\Omega]
		& =2g\rho\Delta(\kappa/2-i\Omega)^{-2}\chi_{\mathrm{m}}\sqrt{2\kappa\eta_{c}\Gamma_{\mathrm{m}}}\nonumber\\
		\rho^{'}&=\left[1+(\kappa/2-i\Omega)^{-2}\Delta\left (\Delta-4g^{2}\chi_{\mathrm{m}}\right )\right]^{-1}.
	\end{align}

	\section{Second-order nonlinear processes}
	
\begin{figure*}[t]
		\centering
		\includegraphics[width=16.5cm]{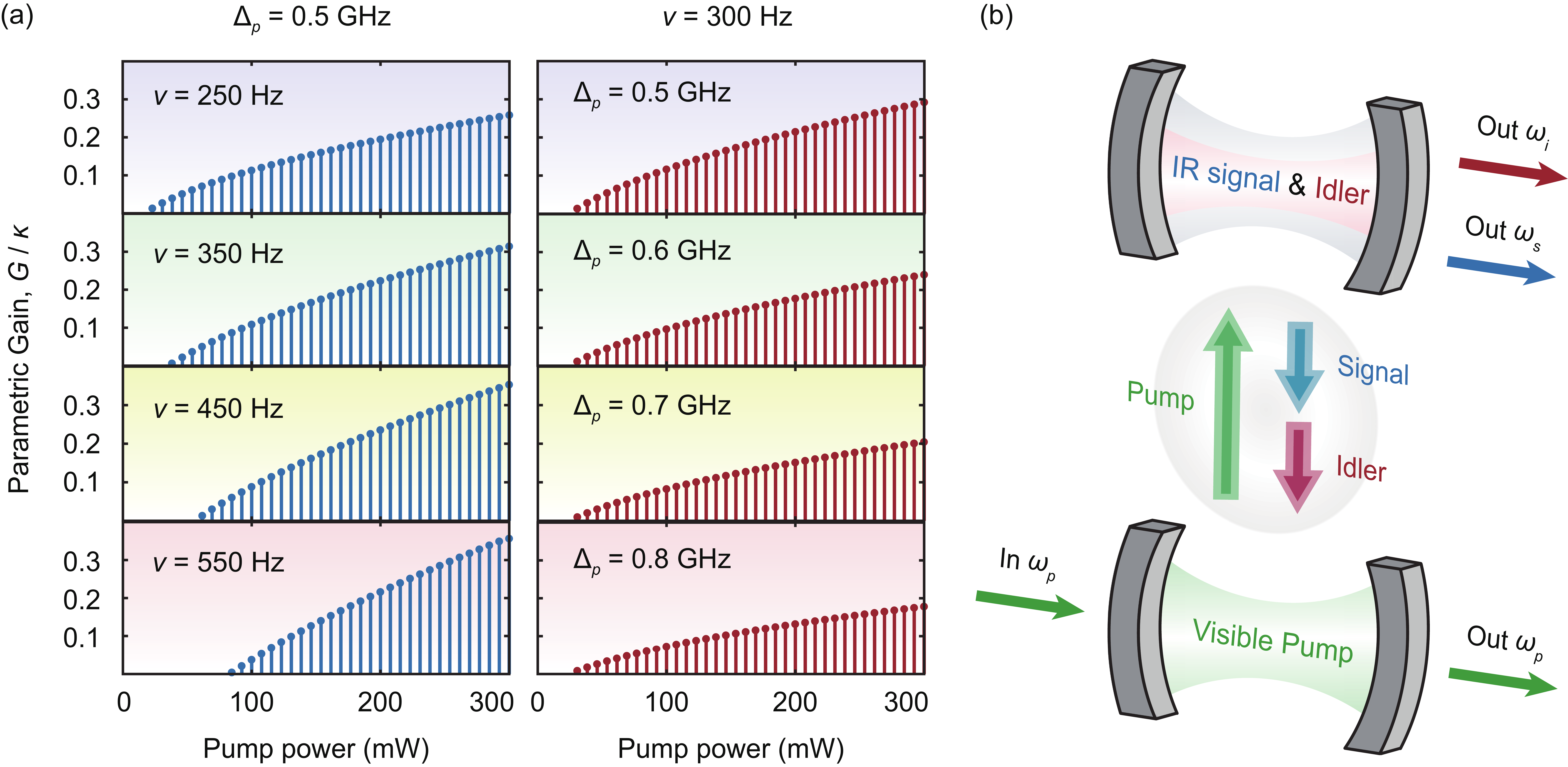}
		\caption{Degenerate OPO process.
			(a) The parametric gain versus the pump power
			for the parametric oscillation process.
			The evolutions are shown at a red detuning
			of the pump field with $\Delta_{p}=\SI{0.5}{\giga\hertz}$
			(left panel)~\cite{SI-Peano2015intracavity} and a
			second-order nonlinearity of $\chi^{(2)}=\SI{300}{\hertz}$
			(right panel)~\cite{SI-Bruch2019opa}. The
			theoretical pump power threshold is tens of
			milliwatts, which is in agreement with the
			recent OPO experiment~\cite{SI-Bruch2019opa}.
			(b) Schematic representation of the OPO model
			using two Fabry–P\'{e}rot cavities.
		}\label{OPO}
	\end{figure*}

	In the case of strong optical drives, the nonlinear gain coefficient is derived from the steady-state
	equations:
	\begin{equation}
		G=\chi^{(2)}\qty(\tau-\sqrt{\tau^{2}+\frac{\Omega_{m}}{8
				g_{0}}-\frac{\abs{{\cal E}_{p}}}{4\chi^{(2)}}}),
	\end{equation}
	where $\tau=\sqrt{\kappa^{2}+4\Delta_{p}^{2}}
	/\qty(16\chi^{(2)})$, $\abs{{\cal E}_{p}}=\sqrt{
		\kappa\eta_{c}P_{p}/\qty(\hbar
		\omega_{p})}$, and $P_{p}$ quantifies the
	pump power for the $\chi^{(2)}$ crystal.
	The nonlinear gain coefficient is enhanced with the
	increase of the  power  of the
	pump laser and the second-order nonlinearity
	[Fig.~\ref{OPO}(a), left panel], following the
	characteristic optical parametric oscillation
	(OPO) power curves~\cite{SI-Bruch2019opa}.
	However, the photons circulating in the cavity
	is reduced when increasing the detuning of the
	pump field, which results in the suppression of
	the nonlinear gain coefficient [Fig.~\ref{OPO}(a),
	right panel]. Figure \ref{OPO}(b)
	schematically illustrates the
	$\chi^{(2)}$ nonlinear process, where the OPO
	model can be treated as two coupled cavities
	with spontaneous parametric
	down-conversion~\cite{SI-Bruch2019opa}.
	The visible pump laser at frequency $\omega_{p}$
	drives the $\chi^{(2)}$ crystal, producing a pair of infrared
	signal and idler lights at frequencies $\omega_{s}$ and
	$\omega_{i}$, which satisfies the energy-matching
	condition $\omega_{p}=\omega_{s}+\omega_{i}$.
	For degenerate OPOs ($\omega_{s}=\omega_{i}
	=\omega_{p}/2$), a single parametric oscillation is
	realized at half the frequency of the pump laser.
	Whereas for non-degenerate cases ($\omega_{s}\neq
	\omega_{i}$), the OPO process is operated at
	two distinct resonances centered about the pump.

	\section{Stability conditions}
	
	The stability or instability of the system
	is determined by the signs of the real parts
	of the eigenvalues of the dynamical evolution
	matrix $\mathbf{M}$. To find the eigenvalues
	$\lambda$, it is necessary to solve the
	characteristic equation $\det(\mathbf{M}-
	\lambda\mathbf{I})=0$, which is reduced to an
	algebraic equation of the 4th degree:
	$\lambda^{4}+\mathrm{M}_{3}
	\lambda^{3}+\mathrm{M}_{2}\lambda^{2}
	+\mathrm{M}_{1}\lambda+\mathrm{M}_{0}=0$.
	Applying the Routh-Hurwtiz method, we obtain
	the necessary and sufficient conditions for
	the system stability:
	\begin{align}\label{R-H}
		0 <\mathrm{M}_{3}, \qquad 0& <\mathrm{M}_{3}
		\mathrm{M}_{2}-\mathrm{M}_{1}, \nonumber\\
		0 <\mathrm{M}_{0}, \qquad 0 &<\mathrm{M}_{3}
		\mathrm{M}_{2}\mathrm{M}_{1}
		-\left(\mathrm{M}_{1}^{2}+\mathrm{M}_{3}^{2}
		\mathrm{M}_{0}\right).
	\end{align}
	
	These conditions allow to determine whether all
	the roots in the characteristic equation have
	negative real parts. Thus, we can use them to
	justify the system stability without solving
	the characteristic equation itself. Herein, we
	focus on the resonance case ($\Delta\approx
	\Delta_{c}=0$), thereby the first three inequalities
	in \eqref{R-H} yield the first two stability
	conditions: $\left.G\middle/\kappa<0.25
	\right.$, $-\pi<\theta<0$.
	To proceed, by exploiting the last inequality in
	\eqref{R-H}, we formulate the stability
	criterion functions $\Theta$ as
	\begin{align}
		\Theta&= \mathrm{M}_{3}\mathrm{M}_{2}
		\mathrm{M}_{1}-\left(\mathrm{M}_{1}^{2}
		+\mathrm{M}_{3}^{2}\mathrm{M}_{0}\right).
	\end{align}
	
	Then, the signs of $\Theta$ provide
	the remaining stability requirements:
	\begin{equation}
		\mathrm{sgn}\left(\Theta\right)=
		\begin{cases}
			1, & \text{implies stability},\\
			\text{otherwise}, & \text{implies instability}.
		\end{cases}
	\end{equation}
	As shown in main text, the
	parameters used in our numerical calculations
	are chosen truly in the stable region.
	In particular, the required signal power can be
	derived from \eqref{steady-eqs}:
	\begin{equation}\label{matched-power}
		P_{c}=\frac{\hbar\Omega_{l}n_{c}}
		{4\kappa\eta_{c}}\qty[\qty(\kappa-4G)^{2}
		+8G\kappa\qty(1-\cos\theta)],
	\end{equation}
	which is tens of microwatts and can be attained
	with accessible experimental conditions~\cite{
		SI-Zhang2018OPA}. In principle, a system tends
	to be sensitive to external perturbations
	in the unstable region. Then, the sensitive
	region in the stable realm locates near the
	dividing line between stability and instability.
	
	\section{Signal-to-noise ratio and the
		optimal variance}
	\begin{figure}[h]
		\centering
		\includegraphics[width=17.5cm]{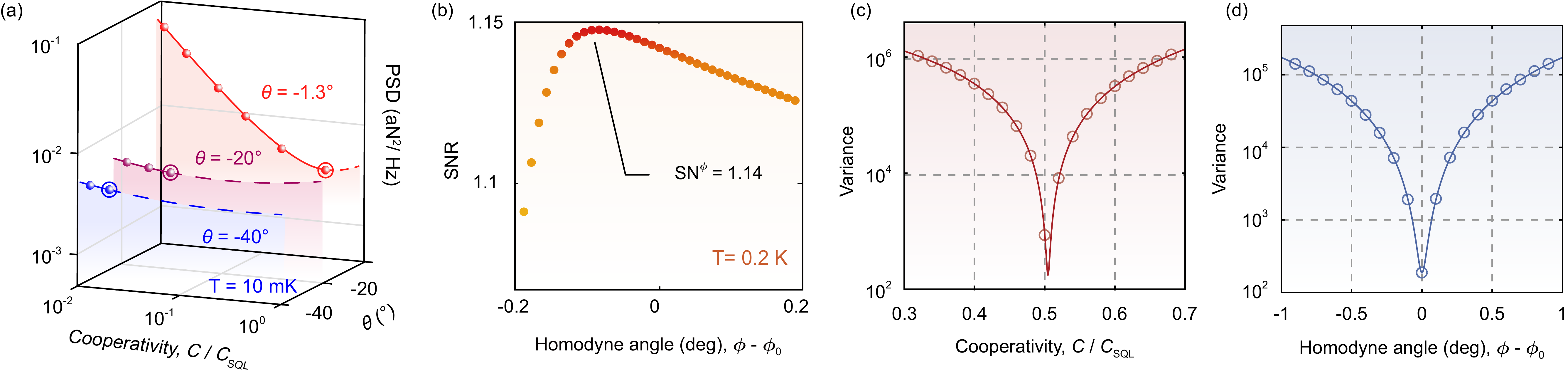}
		\caption{Signal-to-noise ratio (SNR)
			and the optimal variance. (a) Power spectral density (PSD) as a function of  cooperativity and $\theta$. The solid and dashed curves denote the PSD in the stable or unstable region, respectively, and the circles give the minimum PSD in the stable region.  The bath temperature $T =\SI{10}{\milli\kelvin}$.
			(b) SNR for quadratic COM sensors. The reference phase is chosen
			as $\phi_{0}=\SI{-98}{\degree}$. The bath temperature $T = \SI{0.2}{\kelvin}$. (c)-(d)
			The variance of the optimal quadrature.
			The multi-photon cooperativity is
			$C/C_{\mathrm{SQL}}=0.5$, and the signal
			force spectrum is chosens $\qty(
			\SI{0.1}{\atto\newton})^{2}/\si{
				\hertz}$~\cite{PhysRevApplied.15.L021001}.
		}\label{SNR}
	\end{figure}
	
	The performance of the state-of-the-art sensors
	is commonly quantified by the signal-to-noise
	ratio (SNR). In our system, the spectral density
	and the SNR of the signal force are respectively
	estimated by~\cite{SI-sudhir2017quantum}
	\begin{equation}
		\bar{S}_{\mathrm{FF}}^{\mathrm{est},\phi}\qty[\Omega]
		=\bar{S}_{\mathrm{FF}}^{\mathrm{sig}}
		+\bar{S}_{\mathrm{FF}}^{\phi},\qquad
		\operatorname{SN}^{\phi}\qty[\Omega]
		=\frac{\bar{S}_{\mathrm{FF}}^{\mathrm{est},\phi}}
		{\bar{S}_{\mathrm{FF}}^{\mathrm{est},\phi}
			-\bar{S}_{\mathrm{FF}}^{\mathrm{sig}}}.
	\end{equation}
	The spectral density of the apparent force experienced by the oscillator is described as $\bar{S}_{\mathrm{FF}}^{\mathrm{est},\phi}$, and the spectral density of the signal force is described as $\bar{S}_{\mathrm{FF}}^{\mathrm{sig}}$.  As shown in Fig.~\ref{SNR}(a), the
	SNR can reach $1.14$ at the temperature of $0.2$ K.
	
	Figure \ref{SNR}(b)-(c) characterizes
	the variance of the generalized rotated field
	quadrature, which is given by~\cite{
		SI-meng2020mechanical}
	\begin{equation}
		V_{\mathrm{qq}}^{\phi}\qty[\Omega]=\int_{-\infty}
		^{\infty}\Re{\bar{S}_{\mathrm{qq}}^{\phi,\mathrm{out}}
		}\frac{\dd\Omega}{2\pi},
	\end{equation}
	where the optical output spectrum is expressed
	as~\cite{SI-sudhir2017quantum}
	\begin{equation}
		\bar{S}_{\mathrm{qq}}^{\phi,\mathrm{out}}
		\qty[\Omega]=\bar{S}_{\mathrm{qq}}^{\mathrm{out}}
		\left[\Omega\right]\cos^{2}\phi
		+\bar{S}_{\mathrm{pp}}^{\mathrm{out}}
		\left[\Omega\right]\sin^{2}\phi
		+\bar{S}_{\mathrm{pq}}^{\mathrm{out}}
		\left[\Omega\right]\sin\left(2\phi\right).
	\end{equation}
	Such a variance reaches its lowest value when
	choosing proper cooperativity and homodyne angle.
	
	\section{Extended applications to precision
		measurements}
	
	Table~\ref{tab:applications}
	provides a comparison of performance metrics
	for recently reported COM sensors including
	the force sensor described in this
	work.
	
	\begin{table*}[h]
		\centering
		\renewcommand\arraystretch{1.5}
		\begin{tabular}{lllllll}
			\hline
			\hline
			& \makecell[c]{\bf Experiment} &
			& \makecell[c]{\bf Mean phonon} &
			& \makecell[c]{\bf Equivalent force} \\
			\makecell[c]{\bf Sensors} &
			\makecell[c]{\bf (Y/N)} &
			\makecell[c]{\bf Temperature} &
			\makecell[c]{\bf occupations} &
			\makecell[c]{\bf Reported sensitivity} &
			\makecell[c]{\bf sensitivity} &
			\makecell[c]{\bf References}
			\\\hline
			\makecell[c]{Magnetometer}
			& \makecell[c]{Y}
			& \makecell[c]{$\SI{300}{\kelvin}$}
			& \makecell[c]{$\num{1.1e6}$}
			& \makecell[c]{$\qty(400~\si{\nano\tesla})^{2}\si{/\hertz}$}
			& \makecell[c]{$\qty(2.4~\si{\pico\newton})^{2}\si{/\hertz}$}
			& \makecell[c]{\cite{SI-forstner2012cavity}}\\
			\makecell[c]{Magnetometer}
			& \makecell[c]{Y}
			& \makecell[c]{$\SI{300}{\kelvin}$}
			& \makecell[c]{$\num{1.2e6}$}
			& \makecell[c]{$\qty(200~\si{\pico\tesla})^{2}\si{/\hertz}$}
			& \makecell[c]{$\qty(1.2~\si{\femto\newton})^{2}\si{/\hertz}$}
			& \makecell[c]{\cite{SI-forstner2014ultrasensitive}} \\
			\makecell[c]{Magnetometer}
			& \makecell[c]{Y}
			& \makecell[c]{$\SI{300}{\kelvin}$}
			& \makecell[c]{$\sim\num{1.0e6}$}
			& \makecell[c]{$\qty(5~\si{\nano\tesla})^{2}\si{/\hertz}$}
			& \makecell[c]{$\qty(0.75~\si{\nano\newton})^{2}\si{/\hertz}$}
			& \makecell[c]{\cite{SI-Li2018quantum}} \\
			\makecell[c]{Torque sensor}
			& \makecell[c]{Y}
			& \makecell[c]{$\sim\SI{1}{\milli\kelvin}$}
			& \makecell[c]{$\sim 2.8$}
			& \makecell[c]{$\qty(1.3~\si{\zepto\newton.\metre})^{2}\si{/\hertz}$}
			& \makecell[c]{$\qty(0.43~\si{\femto\newton})^{2}\si{/\hertz}$}
			& \makecell[c]{\cite{SI-wu2014dissipative}} \\
			\makecell[c]{Ultrasound sensor}
			& \makecell[c]{Y}
			& \makecell[c]{$\SI{300}{\kelvin}$}
			& \makecell[c]{$\num{1.3e8}$}
			& \makecell[c]{$\qty(8~\si{\micro\pascal})^{2}\si{/\hertz}$}
			& \makecell[c]{$\sim\qty(370~\si{\femto\newton})^{2}\si{/\hertz}$}
			& \makecell[c]{\cite{SI-basiri2019precision}}
			\\\hline
			\makecell[c]{\multirow{4}{*}{This work}}
			& \makecell[c]{\multirow{4}{*}{N}}
			& \makecell[c]{$\SI{300}{\kelvin}$}
			& \makecell[c]{$\num{6.2e6}$}
			& \multicolumn{3}{c}{$\qty(\SI{10.2}{\atto\newton})^{2}
				\si{/\hertz}$} \\
			& & \makecell[c]{$\SI{10}{\kelvin}$}
			& \makecell[c]{$\num{2.1e5}$}
			& \multicolumn{3}{c}{$\qty(\SI{1.86}
				{\atto\newton})^{2}\si{/\hertz}$}\\
			& & \makecell[c]{$\SI{0.2}{\kelvin}$}
			& \makecell[c]{$\num{4.2e3}$}
			& \multicolumn{3}{c}{$\qty(\SI{0.26}
				{\atto\newton})^{2}\si{/\hertz}$}
			\\\hline\hline
		\end{tabular}
		\caption{Extended applications to the
			state-of-the-art COM sensors. The resolution
			of the accelerometers is quantified by
			noise-equivalent acceleration in
			units of $g^{2}\si{/\hertz}$,
			where $1\,g=\SI{9.81}{\metre\per\second^{2}}$.}
		\label{tab:applications}
	\end{table*}

\end{document}